\documentclass[graybox,vecphys]{svmult}

\usepackage{type1cm}        % activate if the above 3 fonts are
\usepackage{makeidx}         % allows index generation
\usepackage{graphicx}        % standard LaTeX graphics tool
\usepackage{multicol}        % used for the two-column index
\usepackage[bottom]{footmisc}% places footnotes at page bottom
\usepackage{newtxtext}       % 
\usepackage{newtxmath}       % selects Times Roman as basic font
\usepackage{ bbold }

\providecommand{\vect}[1]{{\bf{#1}}}

\makeindex             % used for the subject index

\begin{document}

\title*{Current-Induced Dynamics of Chiral Magnetic Structures: Creation, Motion, and Applications}
\titlerunning{Current-Induced Dynamics of Chiral Magnetic Structures}
\author{Jan Masell and Karin Everschor-Sitte}
\institute{Jan Masell \at RIKEN Center for Emergent Matter Science, 2-1 Hirosawa, Wako, Saitama 351-0198, Japan, \email{jan.masell@riken.jp}
\and Karin Everschor-Sitte \at Johannes Gutenberg University, Institute of Physics, Staudingerweg 7, 55128 Mainz, Germany \email{kaeversc@uni-mainz.de}}
\maketitle

\abstract{
Magnetic textures can be manipulated by electric currents via the mechanisms of spin-transfer and spin-orbit-torques. 
We review how these torques can be exploited to create chiral magnetic textures in magnets with broken inversion symmetries, including domain walls and skyrmions. 
These chiral textures can also be moved by (electric) currents and obey very rich dynamics. 
For example, magnetic domain walls feature the famous Walker breakdown, and magnetic whirls are subject to the skyrmion Hall effect, which is rooted in their real-space topology.
These properties led to a variety of potential novel applications which we briefly overview.
}

%%%%%%%%%%%%%%%%%%%%%%%%%%%%%%%%%%%%
%%%%%%%%%%%%%%%%%%%%%%%%%%%%%%%%%%%%
%%%%%%%%%%%%%%%%%%%%%%%%%%%%%%%%%%%%

\section{Introduction}
Magnetic materials have been studied over the centuries for various prospects, in particular yielding the fundamental building blocks in computers that enable us to store tremendous amounts of data and transcending our culture to the age of information technology. 
Permanent magnetism as a key feature in these devices which offers not only fundamentally interesting, but also application-wise impressive and practical phenomena. 
The fact that magnets can be strongly influenced by external magnetic fields is both, a blessing and a curse. 
On the one hand, localized magnetic fields can be used to easily manipulate magnetic states of matter.
On the other hand, magnetic devices are sensitive to invasive, external stimuli. 
Even nowadays, where magnetic mass storages in the form of rotating hard disc drives are steadily replaced by all-electric devices, magnetic recording media still appears throughout our everyday lives. 
Besides their intrinsic advantage of being non-volatile, magnetic recording media have to overcome some challenges such as increasing the speed for reading and writing information and reducing the energy consumption to be competitive with nowadays all-electric information technology. 
For example, in the 1970s, some memories and computers used magnetic bubbles as mobile information carriers which, however, by the 1980s were completely replaced by magnetic hard drives or transistor-based controllers which turned out to be faster and better scalable.
However, since the 1980s, research has unveiled a number of new effects and novel ways to control the static and dynamic properties of magnetic materials. 
These include most importantly (i) chiral magnetic systems and the ability to control their relativistic asymmetric exchange interaction -- the Dzyaloshinskii-Moriya interaction (DMI)~\cite{Dzyaloshinskii1958, Moriya1960} -- and (ii) the ability to generate current-induced spin-torques, in particular spin-transfer torques (STTs)~\cite{Slonczewski1996, Berger1996} and spin-orbit torques (SOTs)~\cite{Miron2011, Liu2012}.  
These spin-torques can be used to manipulate the magnetization directly, providing a new toolbox for potentially more competitive magnetic applications and opening the door to a whole range of interesting new physical phenomena.

\begin{figure}[t]
    \includegraphics[width=0.99\columnwidth]{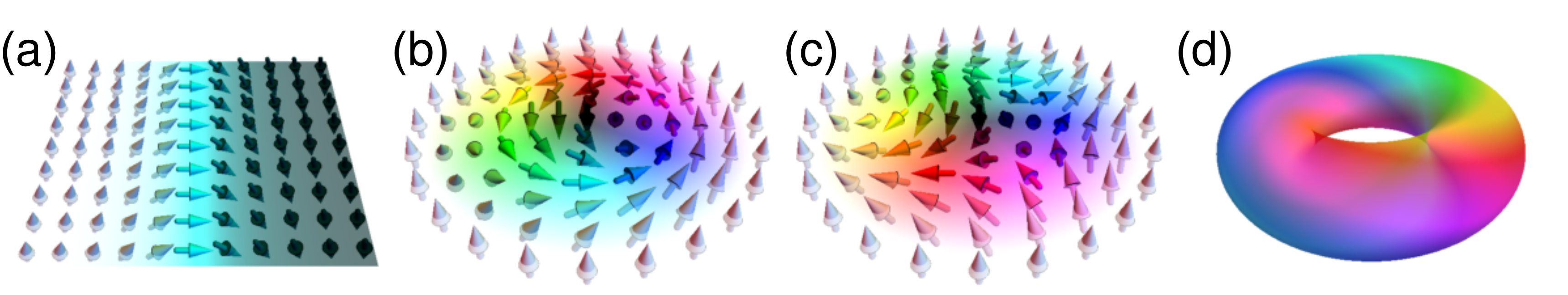}
    \caption{Schematic figures of a (a) N\'eel-type domain wall, (b) Bloch-type skyrmion, (c) antiskymion, and (d) hopfion.
        The color code represents the direction of the normalized local magnetization.
        For the hopfion, we sketch an isosurface of the magnetization, using the same color code as in (a-c).
        }
\label{fig1}       
\end{figure}

This book chapter is intended to serve as an overview over the basic theoretical concepts in the context of chiral magnetic textures and their dynamics, in particular, when subject to spin-torques.
Those spin textures which are stabilized, e.g., in systems with DMI or in systems with strong frustration comprise the well-studied magnetic domain walls,~\cite{Catalan2012} but also the miniaturized versions of magnetic bubbles, i.e., magnetic skyrmions and antiskyrmions,~\cite{Bogdanov1989,Fert2017,Jiang2017, Everschor-Sitte2018, Back2020} and the newly suggested magnetic hopfions.~\cite{Bogolubsky1988, Sutcliffe2017}  
Representatives of such structures are shown in Fig.~\ref{fig1}. 
We first review in Sec.~\ref{sec2:spintorques} the description of magnetic textures within a continuum (micromagnetic) model, discussing their energy functional and their effective dynamic equation -- the Landau-Lifshitz-Gilbert-Slonczewski (LLGS) equation. In this part, we also address the interaction of magnetic textures with electric currents, focusing on the origin and effects of spin-torques.
In Sec.~\ref{sec3:chiralmagnetictextures} we review the most common magnetic textures. 
In Sec.~\ref{sec4:creation}  we address how to create magnetic textures focusing on all-electrical methods.
In Sec.~\ref{sec5:motion} we review the recent progress made in the analysis of the motion of spin textures subject to spin-torques. In particular, we provide a detailed review on one of the most important and yet simple theoretical concepts for the motion of magnetic textures -- the Thiele equation in its generalized form. We demonstrate how to apply it to the dynamics of magnetic textures such as  domain walls, skyrmions, and hopfions. 
Finally, in Sec.~\ref{sec6:applications}, we give a brief overview over the plethora of suggested possible applications for chiral magnetic textures.

%%%%%%%%%%%%%%%%%%%%%%%%%%%%%%%%%%%%
%%%%%%%%%%%%%%%%%%%%%%%%%%%%%%%%%%%%
%%%%%%%%%%%%%%%%%%%%%%%%%%%%%%%%%%%%
\section{Continuum model for the magnetization}
\label{sec2:spintorques}

In this section, we present the continuum description of magnets and their interplay with electric currents, which in a simplified form is known as the micromagnetic model. 

%%%%%%%%%%%%%%%%%%%%%%%%%%%%%%%%%%%%
%%%%%%%%%%%%%%%%%%%%%%%%%%%%%%%%%%%%
\subsection{Magnetization statics}

The static properties of any magnet are well determined by an energy functional whose form depends strongly on the symmetries of the system.
The precise determination of this energy functional in all its components is a very hard task.
For sufficiently simple systems, the spin wave dispersion can be calculated with ab initio methods and then be fitted to a model of localized magnetic moments $\{\vec{S}_i\}$.
Such treatments are very successful in describing magnetism on the atomic scale, which often requires exchange interactions $\vec{S}_i\cdot\vec{S}_j$ beyond nearest neighbors and, potentially, also more exotic interactions between multiple spins.~\cite{Heinze2011}

The magnetization in most chiral ferromagnets is, however, smooth, i.e., it is polarized on the length scale of the atomic lattice and varies only on much larger length scales.
In this limit, the magnetic system can be well described by a phenomenological Ginzburg-Landau theory where an effective energy functional for the magnetization $\vec M$ is derived as a series expansion in powers of $\vec M$ and spatial derivatives $\partial_\alpha$.
Moreover, for temperatures far below the Curie temperature the magnetic system is in an ordered state and the local magnitude of the magnetization corresponds to the saturation magnetization $M_s$.
The resulting energy functional can then be expressed in terms of the normalized magnetization $\vec m = \vec{M}/M_s$ in very general terms as 
\begin{align}
\label{eq:energy}
    E[\vec{m}] = \int\!\!\mathrm{d}\vec{r} \,\,    \big[
    &- B_i \vec{m}_i 
    - K_{ij} \, \vec{m}_i \vec{m}_j 
    - K_{ijkl} \, \vec{m}_i \vec{m}_j \vec{m}_k \vec{m}_l
    \notag \\
    &- D_{ij}^\alpha \,\vec{m}_i \partial_\alpha \vec{m}_j 
    + A_{ij}^{\alpha\beta} \, \partial_\alpha \vec{m}_i \partial_\beta \vec{m}_j
    - Q_{ijk}^{\alpha\beta} \,\vec{m}_i \partial_\alpha \vec{m}_j \partial_\beta \vec{m}_k 
    \\
    &+ A_{ijkl}^{\alpha\beta} \, \vec{m}_i \vec{m}_j \partial_\alpha \vec{m}_k \partial_\beta \vec{m}_l  
    + A_{ij}^{\alpha\beta\gamma\delta} \, \partial_\alpha \partial_\beta \vec{m}_i \partial_\gamma \partial_\delta \vec{m}_j 
    - ... \big] \notag
\end{align} 
where we implicitly sum over all spatial indices $\alpha,\beta$ and magnetization indices $i,j$.
The first term is usually written explicitly as $\vec{B} = \mu_0 M_s ( \frac{1}{2} \vec{H}_\mathrm{d} + \vec{H})$ where $\vec{H}_\mathrm{d}$ is the demagnetizing field and $\vec H$ is the externally applied magnetic field.
All other interaction tensors are material specific and their tensorial structure is determined by the point group symmetry of the system. 
In principle, they can be completely anisotropic and even non-local, similar to the demagnetizing field.
For an effective description of the low energy physics on large length scales, the infinite series in Eq.~\eqref{eq:energy} is restricted to only the most relevant terms.
Higher order interaction processes are usually small which suppresses terms which are higher order in the magnetization.
Higher orders of derivatives, moreover, are suppressed as they become increasingly irrelevant on larger scales.
Other terms, such as the DMI term with $D_{ij}^\alpha$, are only non-vanishing because of the finite spin-orbit coupling, which is usually also small.
Here we list the most common and relevant examples focusing on magnetic systems with their dominant lowest order chiral interaction

\begin{itemize}
\item For a \emph{time-reversal invariant system} all terms with an odd power of $\vec m$ vanish.
\item For an \emph{inversion symmetric system} there is no chiral interaction, i.e.\ the DMI term vanishes, $D_{ij}^\alpha=0$, and so do all terms with an odd number of derivatives.
\item For a \emph{bulk chiral magnet with a cubic unit cell} and a three-fold screw axis in the [111] direction,
like the prototypical chiral magnets MnSi or FeGe, the DMI tensor simplifies to what is denoted as Bloch-type DMI in the literature, i.e., $D_{ij}^\alpha=D \epsilon_{i \alpha j}$.
The exchange interaction becomes $A_{ij}^{\alpha\beta} = A \delta_{ij}\delta_{\alpha\beta} + A'\delta_{ij}\delta_{\alpha\beta}\delta_{i\alpha}$.
The last term proportional to $A'$ reflects an anisotropic exchange coupling which can be present in cubic systems, but for MnSi and FeGe it turns out to be negligible.\cite{Bak1980} 
\item In \emph{thin films or monolayers}, the inversion symmetry along the film normal (e.g., the $\hat z$-direction) is explicitly broken by the sandwich structure of the material or the substrate, but is usually preserved in the other directions. In such a setup the DMI tensor simplifies to what is known as N\'eel-type DMI, i.e.\  $D_{ij}^\alpha=D (\delta_{i\alpha}\delta_{jz} - \delta_{iz}\delta_{j\alpha})$.
The exchange interaction simplifies to $A_{ij}^{\alpha\beta} = A \delta_{ij}\delta_{\alpha\beta} + A^z \delta_{ij}\delta_{iz}\delta_{\alpha\beta}$. 
Besides exchange and DMI, the term that is often relevant in such systems is the uniaxial anisotropy $K_{ij}=K\delta_{ij}\delta_{iz}$. In combination with the demagnetizing field, it can lead to the stabilization of magnetic bubbles.

\item For systems with lower symmetry, the emerging terms and the corresponding tensor entries become more and more complex. We still would like to highlight \emph{systems with $C_{2v}$ symmetry}, where the two-fold rotational symmetry allows not only to realize magnetic skyrmions but also antiskyrmions, \cite{Hoffmann2017} see Fig.~\ref{fig1}. 
In a basis where the $\hat{z}$-axis is the two-fold rotational symmetry and the $\hat{x}$ and $\hat{y}$-axes are defined to be along the two reflection planes of the $C_{2 v}$ point group,~\cite{Birss1964} the exchange parameters are $A_{ij}^{\alpha\beta} \, \partial_\alpha \vec{m}_i \partial_\beta \vec{m}_j = A_i \delta_{ij}\delta_{\alpha\beta}$
 and there are seven independent DMI tensor components given by $D_{xz}^x, \ D_{zx}^x , \  D_{yz}^y,  \  D_{zy}^y, \   D_{zz}^z,  \ D_{xx}^z,  \ \text{and} \  D_{yy}^z$.
 For further interesting systems we refer to Refs.~\cite{Hals2018, Ado2019}.
\end{itemize} 

To summarize, the specific systems determine which magnetic interaction scales are relevant and which magnetic structures can be realized as (meta-)stable states. Over the past century, magnets with strong uniaxial anisotropy have been in the focus of material research, mostly application-oriented.
With the advances made over the past decades, more detailed engineering of the properties of magnetic materials became possible and experimental techniques were developed that enable the observation of magnetic structures on the nanometer scale.
With these new techniques at hand, more exotic materials can be studied where other interactions are dominant and stabilize new forms of magnetic textures.

%%%%%%%%%%%%%%%%%%%%%%%%%%%%%%%%%%%%
%%%%%%%%%%%%%%%%%%%%%%%%%%%%%%%%%%%%
\subsection{Magnetization dynamics in the presence of spin-torques}
The interplay of magnetism and currents is very complex and they mutually influence each other. For example, upon traversing a topologically non-trivial magnetic structure, the electrons pick up a Berry phase~\cite{Berry1984} which leads then leads to a topological Hall 
effect~\cite{Back2020,Neubauer2009, Lee2009, Everschor-Sitte2014} in addition to other Hall contributions such as the anomalous Hall effect.
In this part, we will focus on the effects that an electric current has on the magnetization.

Within the micromagnetic framework, where the local magnitude of the magnetization is constant, the slow and smooth magnetization dynamics can be described effectively within the 
LLGS equation \cite{Slonczewski1996} 
\begin{equation}
d_{t}\vec{m} =
    - \gamma\, \vec{m} \times \vec{B}_{\mathrm{eff}} + \alpha\, \vec{m} \times d_{t}\vec{m} + \tau,
    \label{eq:LLGS}
\end{equation}
where $\gamma$ is the (positive) gyromagnetic ratio, $\alpha$ is the dimensionless Gilbert damping parameter, and 
$\vec{B}_{\mathrm{eff}}=-\delta E[\vec{m}] / (M_{s}\delta\vec{m})$
 is the effective magnetic field due to interactions in the magnetization. $\tau $ represents the current-induced magnetic torques. It comprises STTs as well as SOTs, $\tau = \tau_{\mathrm STT} + \tau_{\mathrm{SOT}}$. 
 Their lowest order terms comprise each a field- and damping-like term\cite{Sampaio2013}
\begin{subequations}
\begin{align}
\tau_{\mathrm{STT}} &=
     - \left(\vec{v}\!_e \cdot \nabla\right) \vec{m} + \beta\, \vec{m}\times( \vec{v}\!_e \cdot \nabla ) \vec{m} \\
    \tau_{\mathrm{SOT}} &=  - \tau_{\mathrm{FL}}\,  \vec{m} \times \vec{\sigma} -  \tau_{\mathrm{DL}}\, \vec{m} \times (\vec{m} \times \vec{\sigma} ),
\label{eq:torques}
\end{align}
\end{subequations} 
where $\vec{v}\!_e = - [P\mu_\mathrm{B}/e M_{s}(1+\beta^{2})]\vec{j}\!_e$ is the effective spin velocity~\cite{Zhang2004} with $\vec{j}\!_e$ the electric current density, $P$ the polarization, $\mu_\mathrm{B}$ the Bohr magneton, and $e>0$ the electron charge. 
$\beta$ is the non-adiabatic damping parameter. 
$\vec{\sigma}$ encodes the spin polarized current: For the typical situation where the SOTs\cite{Miron2011, Liu2012} are generated by the spin Hall effect at an interface between a ferromagnet and a heavy metal, it is $\vec{\sigma}= \hat{\vec n} \times \vec{j}\!_e$ where $\hat{\vec n}$ is the normal direction of the interface between the materials. The strengths $\tau_{\mathrm{FL}}$ and $\tau_{\mathrm{DL}}$ for the field-like and damping-like terms are material dependent.

\section{Magnetic solitons}
\label{sec3:chiralmagnetictextures}

In this part we review the most common magnetic structures focusing on chiral solitons, shown in Fig.~\ref{fig1}.

%%%%%%%%%%%%%%%%%%%%%%%%%%%%%%%
% Magnetic domain walls
%%%%%%%%%%%%%%%%%%%%%%%%%%%%%%%
\runinhead{Magnetic domain walls} are rather ubiquitous one-dimensional textures that connect two distinctly polarized phases. 
The reason for this is that they do not require any particular stabilization mechanisms; the two distinct ferromagnetic ordered phases can be fixed by the boundary conditions. 
Therefore, magnetic domain walls have been observed and studied already long ago and can be found in many different samples with different properties. 
By continuation in further dimensions, domain walls can also be hosted in two-dimensional or three-dimensional systems.  
For example, in symmetric thin films, Bloch- or N\'eel-type domain walls can be stabilized. 
Their helicity is determined by magnetostatic interactions and therefore depends on the film geometry. 
For very thin films, N\'eel type domain walls are formed, where the magnetization winds from one out-of-plane polarized state to the oppositely polarized state in the plane spanned by the out-of-plane state and the direction of rotation, as shown in Fig.~\ref{fig1}.
For thicker films mainly Bloch type domain walls are realized. 
In two-dimensional systems, domain walls can be effectively described as strings~\cite{Malozemoff1973, Rodrigues2018} and closing this string leads to structures that are called magnetic bubbles. 
Furthermore, domain walls can obey localized defects for example in the version of Bloch lines, i.e., localized windings in the domain wall where the helicity switches from one Bloch handedness to the other handedness.

%%%%%%%%%%%%%%%%%%%%%%%%%%%%%%%
% Magnetic skyrmions
%%%%%%%%%%%%%%%%%%%%%%%%%%%%%%%
\runinhead{Magnetic skyrmions} are localized whirls in two dimensions which can be viewed as a closed magnetic domain wall, embedded as defects in a surrounding background phase or they can be ordered in a lattice. 
In three-dimensional systems, skyrmions form extended strings. 
Skyrmions received lots of attention in particular due to their non-trivial real-space topology. 
The two-dimensional winding number for skyrmions (located in the $xy$-plane)
\begin{equation}
\mathcal{Q} = \frac{1}{4\pi}\int_\Omega \mathrm{d}\vec r \,\, \vect{m}\cdot(\partial_{x}\vec{m}\times \partial_{y}\vec{m}) 
= \frac{1}{4\pi } \int_\Omega \mathrm{d}\vec r \, F_z \in \mathbb{Z}
\label{eq:skyrmionnumberQ}
\end{equation}
evaluates to $\mathcal{Q}=-1$ for the skyrmion and to $\mathcal{Q}=+1$ for the antiskyrmion shown in Fig.~\ref{fig1}, when integrating over the open area $\Omega$ of the skyrmion. 
Note that $\mathcal{Q}$ only evaluates to an integer if $\Omega$ is a closed surface, i.e. $\partial\Omega=0$, which can, however, be mapped to an open area $\Omega$ with a topologically trivial boundary $\partial\Omega$.
In the second equality we have introduced the solenoidal gyro-vector field $\vec{F}$ as 
\begin{equation}
\label{eq:Falpha}
F_{\alpha} = \frac{1}{2} \epsilon_{\alpha \beta \gamma} \vect{m}\cdot \left(\frac{\vec{m}}{\partial r_{\beta}}\times \frac{\vec{m}}{\partial r_{\gamma}} \right) 
\end{equation}
with $\epsilon_{\alpha \beta \gamma}$ being the Levi-Civita symbol. 
While for the skyrmion only one component of this vector field is important, the topological index in 3D -- the Hopf invariant -- involves all components, see below.

Note that from a topological point of view, skyrmions and the earlier studied magnetic bubbles are equivalent. 
Even, given the various systems where such magnetic whirl-like textures with a winding number of $\pm1$ occur, a clear definition and full disentanglement might not be possible. Here, we will refer to magnetic bubbles when the domain wall width of the topological whirl-like structure is small compared to its the center area and to a skyrmion otherwise.
While a strict differentiation between the two is not possible, the static and dynamic properties of magnetic whirls do depend on their detailed energy scales, and can be very different.
In particular, skyrmions are typically smaller and more stable such that they are potentially interesting for future technological applications, see Sec.~\ref{sec6:applications}.

Roughly speaking, skyrmions occur in systems with competing interactions, of which some favor the alignment of magnetic moments, and others prefer their twisting. In most systems, however, it is a more complicated interplay that finally stabilizes the topological magnetic whirls.
Experimentally, skyrmions have first been observed in bulk crystals with broken inversion symmetry as a result of a competition between a uniform stiffness $A$, DMI strength $D$, an applied magnetic field $B$ and strong thermal fluctuations at temperatures slightly below the critical temperature.~\cite{Muhlbauer2009} 
By now several other systems have been identified to host skyrmions, revealing alternative stabilization mechanisms such as spatial confinement and frustrated exchange, e.g., via RKKY.~\cite{Bezvershenko2018, Hirschberger2019}
Moreover, materials have been tailored to exhibit a strong interfacial DMI to host skyrmions at room temperature.~\cite{Boulle2016,Moreau-Luchaire2016}  
For an overview of different material systems we refer to Ref.~\cite{Everschor-Sitte2018}. 

Typically, when discussing magnetic skyrmions, it is assumed that these are whirls in an out-of-plane polarized background. 
However, just as domain walls, skyrmions can be hosted by in-plane polarized backgrounds~\cite{Gobel2019a, Zarzuela2020} %Zhang2015l is broken.
 or even more complex background phases such as conical backgrounds in 3d,~\cite{Leonov2016} 
 or embedded inside a helical phase.~\cite{Muller2017}
While skyrmions are effectively two-dimensional structures, there is an ongoing search to find three-dimensional magnetic solitons.  

A bit in the middle of two or three-dimensional structures are \emph{magnetic bobbers},~\cite{Rybakov2015} which, for example, occur in extended films. 
They look like a skyrmion on the top surface and then turn into a Bloch point within the material. Chiral bobbers are metastable states which are stabilized by the interplay of DMI and the boundary condition. The DMI induces a repulsive force between the skyrmion at the surface and the  the Bloch point, wherefore the remaining skyrmion string is not expelled from the material.
Similar surface effects have been know to occur due to demagnetization effects.~\cite{Malozemoff1979}

%%%%%%%%%%%%%%%%%%%%%%%%%%%%%%%
% Magnetic hopfions
%%%%%%%%%%%%%%%%%%%%%%%%%%%%%%%
\runinhead{Magnetic hopfions} are three-dimensional topological objects which, similar to the relation between skyrmions and domain walls, can be viewed as a closed skyrmion string, see Fig.~\ref{fig1}. 
They can be characterized by the Hopf index $H$, which can be calculated by the White-head formula~\cite{Whitehead1947} 
\begin{equation}
H= - \frac{1}{2 \pi^2} \int_{\mathbb{R}^3} d\vec r \, (\vec F \cdot \vec A)
\end{equation}
with the vector field $\vec F$ defined above in Eq.~\eqref{eq:Falpha} and $\vec A$ being an appropriate vector potential $\nabla \times \vec A = \vec F$.
Though, they have been predicted to occur magnetic systems,~\cite{Bogolubsky1988, Sutcliffe2017}
so far, hopfions have not been observed experimentally neither in magnetic systems nor in solids at all.~\cite{Rybakov2019,Gobel2020}, 

\runinhead{Other topological magnetic textures}
apart from the above mentioned ones, are predicted including those which have a more complex order parameter than just the normalized magnetization. 
Several of them have not yet been observed experimentally. 
However, the vast progress in recent years, allowing to engineer coupling strengths and image magnetizations in more and more detail, might reveal more exotic states in the future.

%%%%%%%%%%%%%%%%%%%%%%%%%%%%%%%
% Creation of Magnetic Solitons
%%%%%%%%%%%%%%%%%%%%%%%%%%%%%%%
\section{Creation of Magnetic Solitons}
\label{sec4:creation}

\begin{figure}[tb]
\includegraphics[width=0.99\columnwidth]{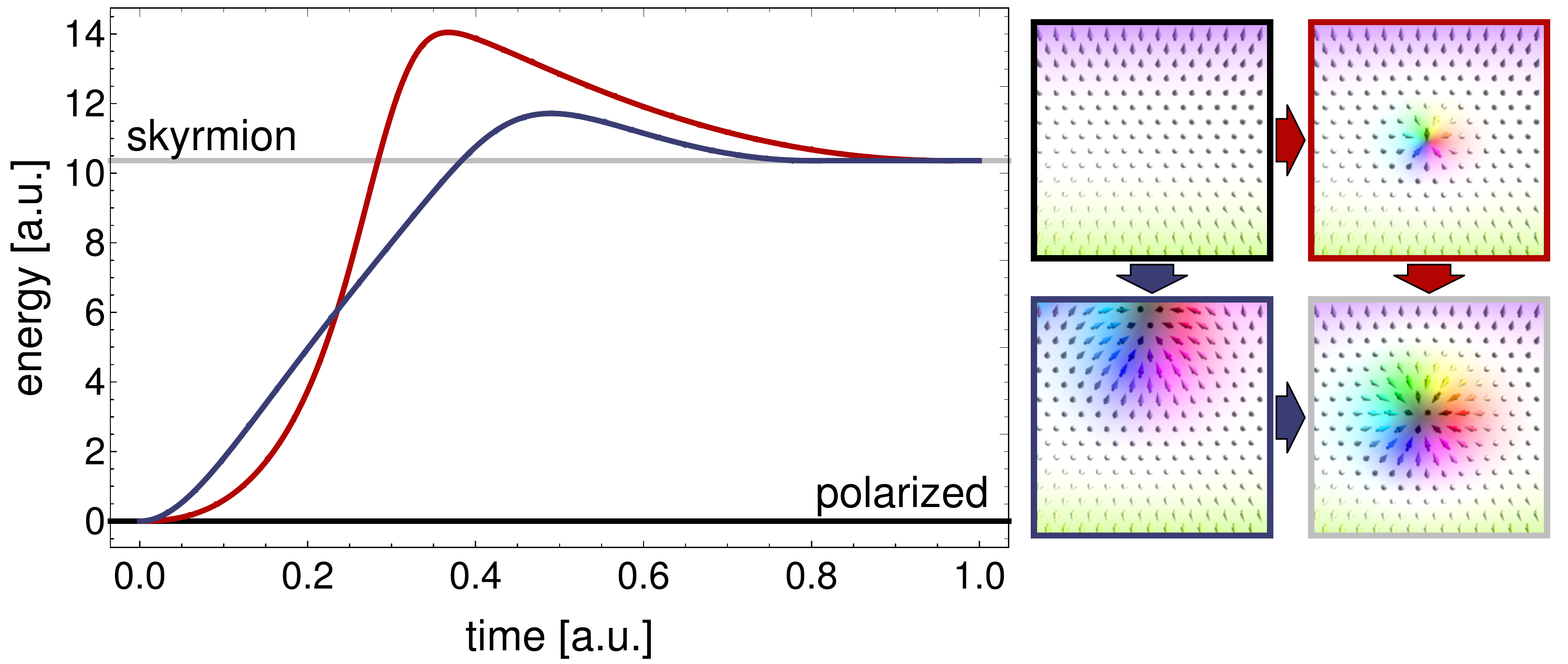}   
\caption{
    Energetics of skyrmion creation in a finite size system with interfacial DMI. 
    The energy barrier depends on details of the interactions but also on the creation process which can, for example, involve a continuous change of the winding number via the edge of the system (blue path) or a discontinuous change via the creation of a skyrmion in the bulk (red path).
    }
\label{fig2energy}      
\end{figure}

In this section we discuss, from a theoretical point of view, how to create magnetic solitons in different dimensions.
These solitons comprise domain walls and skyrmions,
see Sec.~\ref{sec3:chiralmagnetictextures}, and can be introduced into a given magnetic background, such as the ferromagnetic or helical state. 
Before discussing specific properties of different creation mechanisms, we first comment on a few very generic principles.

A soliton is stable and does not decay into magnons if it is protected by a (free) energy barrier. 
Vice versa, the creation of a soliton is also associated with an energy barrier, otherwise the solitons would just spontaneously proliferate and trigger a phase transition. 
As an example, Fig.~\ref{fig2energy} shows two possible mechanisms for the creation/annihilation of a skyrmion in a finite-size two-dimensional system.
On the blue path, the magnetization twists at the edge and a skyrmion enters from outside the sample, while on the red path the skyrmion emerges between lattice points within the sample and then grows.
In either case, the energy (as function of time in arbitrary units) has to rise above the bare energy difference between the initial and the final state but the absolute height of the barrier depends on how the soliton is introduced.
Furthermore, introducing a soliton into the system requires to ``twist'' some parts of the current magnetization state, i.e., exerting local torques on the magnetization structure which are also very different for the two distinct paths shown in Fig.~\ref{fig2energy}.
Thus, the different creation mechanisms can be classified by the effective dimensionality of the magnetic soliton, its embedding background,
and the origin of the acting torques. 
While creating one and two-dimensional textures is explored quite well, the controlled creation of three-dimensional magnetic structures is subject to current and future research.

\subsection{Creation of one-dimensional solitons}
\begin{figure}[tb]
\includegraphics[width=0.99\columnwidth]{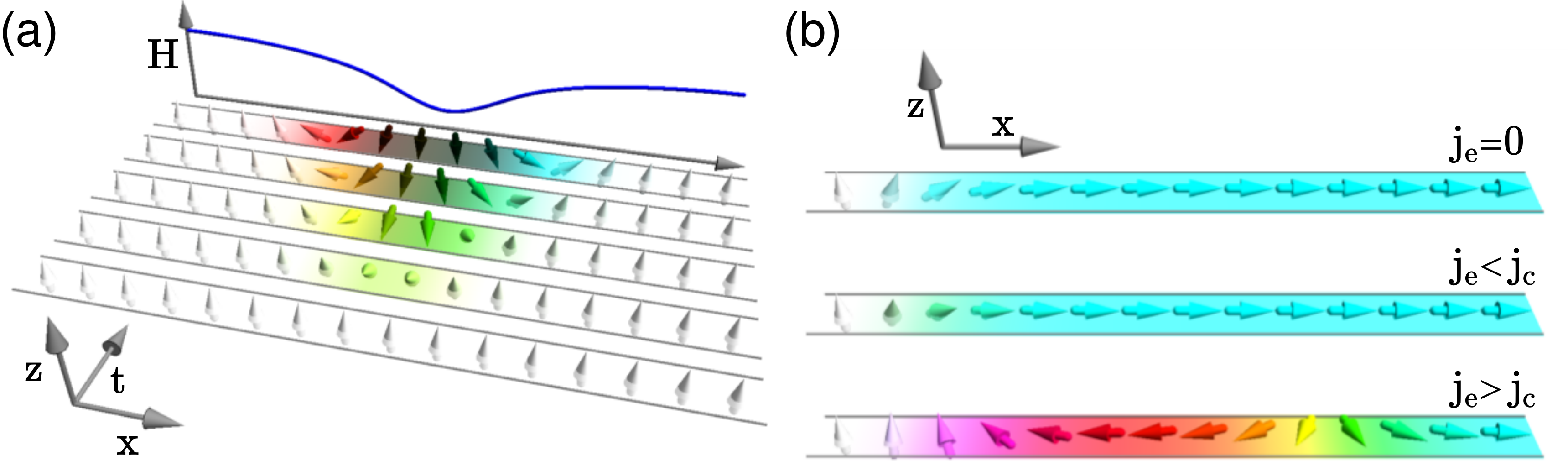} 
\caption{
    Possible mechanisms to create magnetic domain walls.
    (a) Pairwise creation in the middle of a nanowire by a local magnetic field $H$ or spin-currents (not shown).
    (b) Insertion of individual domain walls at the end of the wire via the interplay of spin-torques and an inhomogeneity (white spin fixed e.g.\ by strong perpendicular anisotropy).
}
\label{fig3domainwall}    
\end{figure}

A magnetic domain wall is an (effectively) \emph{one-dimensional} magnetic soliton which usually connects two oppositely polarized phases, see Sec.~\ref{sec3:chiralmagnetictextures}. 
Within a nanowire with uniform magnetization, domain walls can only be created pairwise, as an odd number of domain walls naturally leads to opposite background orientations on both ends. 
To create such a pair of domain walls, one somehow has to locally flip the orientation of the magnetization. 
The most naive way is to locally apply a magnetic field in the desired direction, see Fig.~\ref{fig3domainwall}(a). 
An alternative is to switch the magnetization by means of locally applied spin-currents.

Single domain walls can be created at the edge of the sample. 
One can employ similar techniques as mentioned above, but at the edge the restrictive condition of having the same ferromagnetic state on both sides of the created magnetic texture does not apply.
Alternatively, one can utilize magnetic inhomogeneities in the sample which effectively act as the edge of a smaller subsample.
When an inhomogeneity alters the local magnetization direction, the generation of domain walls is not necessarily pairwise, see Fig.~\ref{fig3domainwall}(b).
One idea is to exploit that the magnetic profile around an inhomogeneity is twisted and, therefore, spin-torques can act on this part by both, further twisting and pulling on the magnetic texture.~\cite{Sitte2016} 
Increasing the applied current will enhance the twisting until a domain wall structure is built, that at the critical current density $j_e^c$ rips off and travels along the system.
Such a creation mechanism also works in a minimal model consisting of exchange and anisotropy interaction and basic STTs.~\cite{Sitte2016}
In this setup, domain walls are created periodically with a period $T$ that depends on the applied current strengths $j_e$, or respectively on the effective spin velocity $v_e$ as  
\begin{equation}
T \sim (v_e-v_e^{c})^{-1/2}\sim (j_e-j_e^c)^{-1/2},
\end{equation}
where the exponent is independent of the microscopic details.
This universal behavior of the shedding period $T$ can be proven by explicitly solving for the magnetic profile and its shedding period in the one-dimensional model including only exchange and anisotropy interactions. 
Furthermore, it is valid for a large class of magnetic systems independent of the details of the microscopic Hamiltonian, including the applicability for higher dimensions.~\cite{Everschor-Sitte2016}  
The required assumptions are 
(i) presuming a translationally invariant model away from the inhomogeneity and 
(ii) neglecting non-adiabatic spin-torque terms.
The argument for the universal exponent in the shedding period is based on combining three ingredients: 
\begin{enumerate}
\item[(1)] the postulate of a critical current density $j_e^c$ above which there will be no statically stable solution and the created magnetic texture rips off the inhomogeneity, 
\item[(2)] the behavior of the magnetic structure in the ``just still static limit'', i.e., for $j_e \lesssim j_e^c$ and 
\item[(3)] the ``just dynamic limit'', i.e.\ for $j_e \gtrsim j_e^{c}$.
\end{enumerate}
For the last two,  one employs that the magnetic profile at the critical point will not differ too much in these two limits. 
The main influence on the magnetic structure will be a (time-dependent) shift in the position $x_0$ where the structure is centered in combination with a mild perturbation on the profile. 
Solving the LLGS equation in these two limits, yields for the ``just still static'' limit the relation $j_e^{c} - j_e^s \sim x_{0}^{2}$ and for the ``just dynamic limit'' $\partial_{t} x_{0} = j_e^d - j_e^s$, where $j_e^s$ is the current strength in the just still static limit and $j_e^d$ in the just dynamic limit. 
These relations are the simplest, that satisfy the expected behavior:  
(i) the velocity of the domain wall depends linearly on the current strength beyond the threshold value and 
(ii) inverting the direction of the current should, in principle, create the domain wall structure in the opposite direction. 
Eliminating $j_e^s$ allows to calculate the period of the magnetic texture formation $T \sim (j_e - j_e^{c})^{-1/2}$ and thus explains the universal dependence.

Note that this universal behavior holds independent of the dimension, provided the above mentioned assumptions are satisfied.
In dimensions higher than one the precise shape of the created magnetic texture cannot be calculated analytically. 
Based on topology, one can, however, conclude that the winding number during the production process must be conserved, opening up the possibility to shed more complex topological structures and their anti-particles.

\subsection{Creation of two-dimensional solitons}

\begin{figure}
\includegraphics[width=0.99\columnwidth]{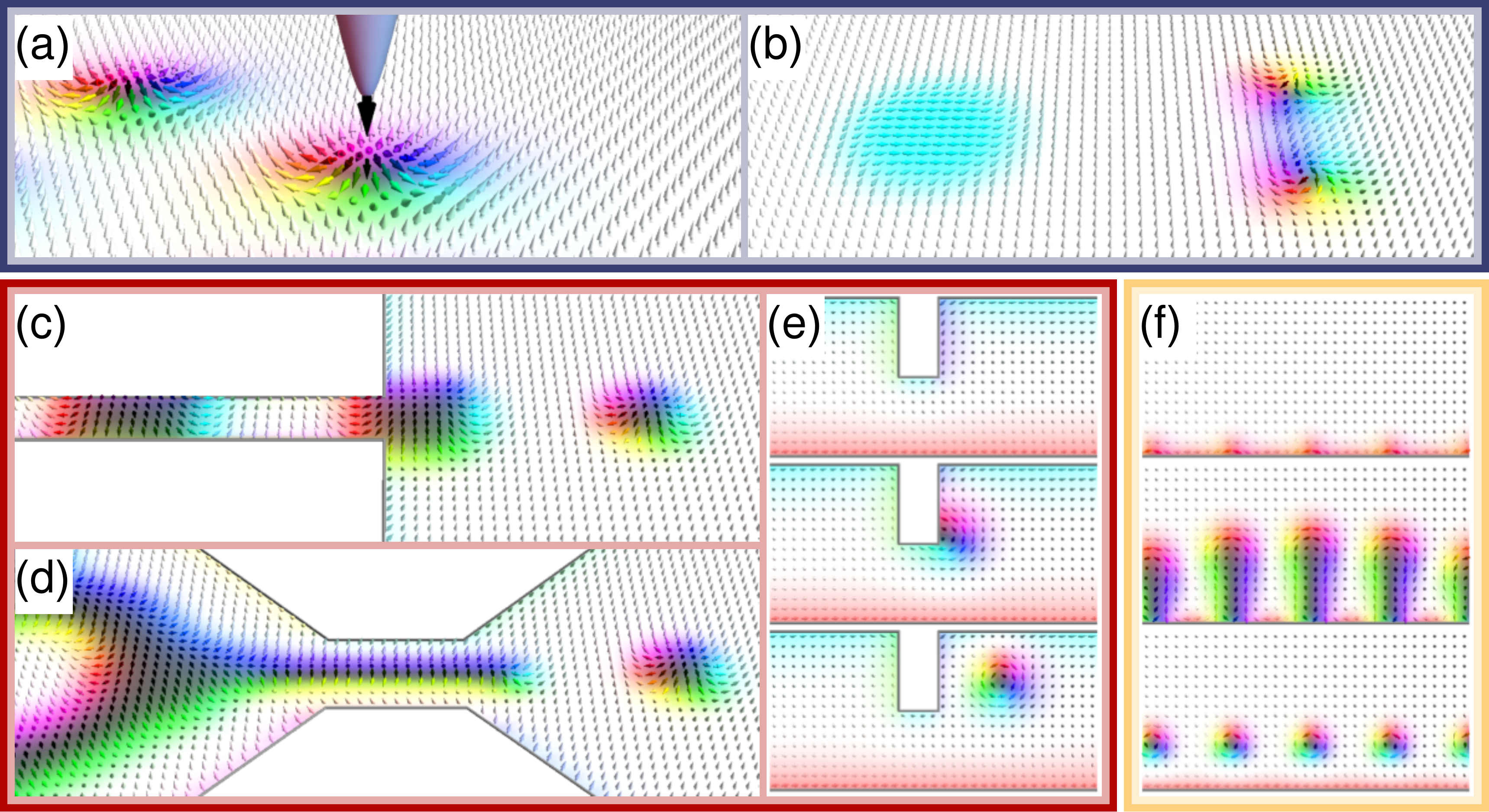}    
\caption{Possible mechanisms to create magnetic skyrmions in ferromagnetic background
by creating them within the sample (blue box), at the boundary (orange box) or via an engineered geometry (red box): 
(a) writing them locally, e.g. with spin-polarized electric currents,
(b) creation of skyrmion-antiskyrmion pairs due to the interplay of magnetic inhomogeneities and spin-torques.
(c) a train of skyrmions is created due to a particular sequence of applied magnetic fields. 
(d) a skyrmion is created at a notch via spin currents.
(e) current-driven domain wall pairs are fusing to skyrmions, and
(f) at the end of a constriction domain walls are chopped off to form skyrmions. 
}
\label{fig4skyrmion}     
\end{figure}

Examples of (effectively) \emph{two-dimensional} magnetic solitons are skyrmions and antiskyrmions, see Sec.~\ref{sec3:chiralmagnetictextures}. 
To create skyrmions numerous methods exist, see for example Ref.~\cite{Everschor-Sitte2018} for an overview.
Similar to domain walls, their creation mechanisms can be categorized by 
(i) being created within the sample,
(ii) at the boundary, or 
(iii) because of a specialized geometry, see Fig.~\ref{fig4skyrmion}. 

To create skyrmions within the sample in a ferromagnetic background one has to invert the magnetization in a small region. 
This can be done for example by local magnetic fields,~\cite{Flovik2017, Zhang2018b}, by local spin currents flowing perpendicular to the material,~\cite{Sampaio2013, Durrenfeld2017} by electric fields induced e.g.\ by spin-polarized STM~\cite{Romming2013, Hsu2017}, by effective local heating~\cite{Koshibae2014} or spontaneously by fluctuations.~\cite{Rendell-Bhatti2019}
Furthermore, one can generate skyrmions dynamically by means of the interplay of spin-currents and some inhomogeneity or defect, as indicated above when discussing domain wall production. 
Increasing the spin-current density above a critical values allows to produce skyrmion-antiskyrmion pairs dynamically by means of STTs.~\cite{Sitte2016, Everschor-Sitte2016, Stier2017, Leonov2016c} 
While the creation mechanism itself is independent of micromagnetic details and, in principle no twisting-like interactions such as DMI are required, in the subsequent dynamics, only the (meta-)stable solutions will continue to exist. 
For example, in a material with Bloch DMI, the antiskyrmion will annihilate and only a Bloch skyrmion will remain. 
Similarly, skyrmions can be created by SOTs.~\cite{Hrabec2016, Legrand2017, Buttner2017a}

An alternative is to create skyrmions via exploiting the tailored geometry of the material, see Fig.~\ref{fig4skyrmion} for different options. 
For example, one can convert a domain wall pair into a skyrmion,~\cite{Zhou2014} or one can generate skyrmions through what has become known as ``blowing bubble'' technique,~\cite{Jiang2015a} where a worm domain is sent through constrictions and ``chopped'' into pieces, i.e.\ skyrmions, by means of the diverging current upon leaving the constriction. 
Or skyrmions can be produced at a notch.~\cite{Iwasaki2013a}
The latter technique leads over to the another principal option, i.e.\ to create skyrmions at the boundaries of a sample. 
Here the effect of the chiral surface states are helpful in pre-twisting the magnetic configurations. 
By means of a properly chosen protocol of an applied field strength one can even generate a whole train of skyrmions at the boundary.~\cite{Muller2016a}

While in a ferromagnetic background single magnetic skyrmions are (meta-)stable states, magnetic skyrmions can be the ground state of a chiral magnet in the form of skyrmion lattices under certain conditions.~\cite{Muhlbauer2009} 
To switch from the competing stripe domain phase into the skyrmion phase several methods exist, including triggering the magnetic material by means of AC field excitations.~\cite{Woo2016}

%%%%%%%%%%%%%%%%%%%%%%%%%%%%%%%%%%%%
%%%%%%%%%%%%%%%%%%%%%%%%%%%%%%%%%%%%
%%%%%%%%%%%%%%%%%%%%%%%%%%%%%%%%%%%%
\section{Motion of Magnetic Solitons}
\label{sec5:motion}

The micromagnetic dynamics of the magnetization are mainly governed by the LLGS equation, Eq.~\eqref{eq:LLGS}.
This equation describes the local precession and relaxation of the magnetization, formulated in terms of a local effective magnetic field which accounts for the interaction of the magnetization with itself and its environment and, moreover, additional torques due to current-induced STTs and/or SOTs.
In general, these non-linear dynamics lead to a complicated dynamical behavior which can even trigger the creation of magnetic solitons as described in Sec.~\ref{sec4:creation} and can usually only be solved numerically.

However, once the solitons are created, they are influenced by the applied spin-torques and other external forces, e.g., due to field gradients.
The reaction of the magnetization is most strongly expressed in the low energy degrees of freedom. 
An effective and potentially more efficient description of the soliton dynamics can therefore be formulated by taking only a few collective coordinates into account.
We will review the derivation of these effective equations of motion, known as (generalized) \emph{Thiele equations},\footnote{Note that the original equation that Thiele derived in his seminal works~\cite{Thiele1973,Thiele1974} refers to the steady-state motion of domain walls. By now the concept that Thiele used to obtain his equation of motion for the domain wall has been generalized for any structure described by a finite set of collective coordinates.} in the following.
We then show examples for their application when we use them as the starting point for the discussion of the dynamics of current-driven magnetic solitons.

%%%%%%%%%%%%%%%%%%%%%%%%%%%%%%%%%%%%
%%%%%%%%%%%%%%%%%%%%%%%%%%%%%%%%%%%%
\subsection{A collective coordinate approximation: Thiele equations of motion}

The main step to obtain the Thiele equations for a given magnetic structure is to project the LLGS equation onto the corresponding collective coordinates. 
This said, the first step is to select suitable collective coordinates for a given magnetic structure. 
In principle, these collective coordinates can represent any property of the quasiparticle. 
To achieve an accurate description of the system with only a few coordinates it makes sense to choose coordinates which are related to zero modes or low energy modes as these are most easily activated, and thus most relevant for the low-energy physics of the system. 
A suitable choice of coordinates should therefore depend on the symmetries of the entire setup: the quasi-particle itself, the energy landscape, and the acting spin-torques. 

To give an example for an appropriate collective coordinate, let us consider the standard assumption of the standard Thiele approach, i.e., a translational invariant model with a \emph{rigid} magnetic texture. 
This means that the magnetic texture does not change its shape when driven by an electric current. 
In this situation, the position of the quasi-particle $\vec{R}(t)$ is a proper collective coordinate (or more generally, \emph{any} position of the rigid magnetic structure) and the magnetization behaves as $\vec{m}(\vec{r},t) = \vec{m}(\vec{r}-\vec{R}(t),0)$.

For the derivation of the generalized Thiele equations, suppose that the time-dependence of the magnetic texture $\vec{m}(\vec{r},t)$ is described by $N$ collective coordinates $\vec{q}(t)=\{q_i(t)\}_{i=1,...,N}$. 
We first isolate the expression for the effective magnetic field $\vec{B}_\mathrm{eff}$ by multiplying the LLGS equation, Eq.~\eqref{eq:LLGS}, by $\vec m \times$ from the left.\footnote{We exploit that the magnetization is a normalized vector field with $|\vec m (\vec r)|=1$. Thus, $\vec m \perp \partial_i \vec m$ and $\vec m \times\vec m \times \partial_i \vec m = -\partial_i \vec m$ for all coordinates $i=x,y,z,t$ and, moreover, $\vec m \perp \vec{B}_\mathrm{eff}$. The latter is always achieved by adding a term $\lambda(\vec{r}) (1-m^2)=0$ to the energy functional which does not change the energy but cancels all components of $\vec{B}_\mathrm{eff}$ that are parallel to $\vec m(\vec r)$.}
Next, we project the LLGS equation onto the translational mode $\frac{\mathrm{d}\vec m}{\mathrm{d} \vec{q}_i}$ of the i-th collective coordinate $\vec{q}_i$, where the projection $\mathcal{P}(\vec{q}_i)$ is implemented by the scalar product 
$\mathcal{P}(\vec{q}_i) = \langle \frac{\mathrm{d}\vec m}{\mathrm{d} \vec{q}_i} |\, . \,  \rangle 
= \int \mathrm{d}\vec{r} ( \frac{\mathrm{d}\vec m}{\mathrm{d} \vec{q}_i} \cdot .\, )$.
Moreover, we explicitly use that all time-dependence is now expressed in the collective coordinates to replace 
$d_t \vec m = \sum_{j=1}^N\dot{\vec{q}}_j \frac{\mathrm{d}\vec m}{\mathrm{d}\vec{q}_j}$ where $\dot{\vec{q}}_j = d_t \vec{q}_j$. 
A compact form of the $i=1,...,N$ Thiele equations for an arbitrary magnetic texture with both STTs and SOTs then reads
\begin{equation}
    F_i(\vec q) = 
    \mathcal{G}_{ij} \dot{\vec{q}}_j 
    + \alpha \mathcal{D}_{ij} \dot{\vec{q}}_j 
    + \mathcal{G}_{i\mu}^\mathrm{STT} \vec{v}_{e,\mu} 
    + \beta \mathcal{D}_{i\mu}^\mathrm{STT} \vec{v}_{e,\mu} 
    + \tau_{\mathrm{FL}} \mathcal{G}_{i\mu}^\mathrm{SOT} \vec{\sigma}_\mu
    + \tau_{\mathrm{DL}} \mathcal{D}_{i\mu}^\mathrm{SOT} \vec{\sigma}_\mu
\label{eq:Thiele:general} 
\end{equation} 
with implicit summation over both, the collective coordinates $j=1,...,N$ and the spatial dimensions $\mu=x,y,z$.
The projection of the effective magnetic field 
$\vec{B}_{\mathrm{eff}}=-\delta E[\vec{m}] / (M_{s}\delta\vec{m})$
can be interpreted as a force 
\begin{equation}
F_i(\vec q) = -\frac{\gamma}{M_{s}} \frac{\mathrm{d}E[\vec{q}]}{\mathrm{d}q_i} 
= - \frac{\gamma}{M_{s}} \int \frac{\mathrm{d}\vec m}{\mathrm{d} q_i} \cdot \frac{\delta E}{\delta \vec{m}} \mathrm{d}\vec r
\end{equation}
acting on the $i$-th collective coordinate $\vec{q}_i$.
Moreover, Eq.~\eqref{eq:Thiele:general} is implicitly non-linear as, in general, all the matrices on the right hand side depend on $\vec{q}(t)$ and explicitly read
\begin{align}
 \mathcal{G}_{ij} &= -\int \vec m \cdot \left( \frac{\mathrm{d}\vec m}{\mathrm{d} q_i} \times  \frac{\mathrm{d}\vec m}{\mathrm{d} q_j} \right)   \mathrm{d}\vec{r},
&\mathcal{D}_{ij} &= \int  \left(\frac{\mathrm{d}\vec m}{\mathrm{d} q_i} \cdot  \frac{\mathrm{d}\vec m}{\mathrm{d} q_j} \right)   \mathrm{d}\vec{r},  \nonumber\\
\mathcal{G}_{i\mu}^\mathrm{STT} &= -\int \vec m \cdot \left( \frac{\mathrm{d}\vec m}{\mathrm{d} q_i} \times  \frac{\mathrm{d}\vec m}{\mathrm{d} x_\mu} \right)   \mathrm{d}\vec{r},  
&\mathcal{D}_{i\mu}^\mathrm{STT} &= \int  \left(\frac{\mathrm{d}\vec m}{\mathrm{d} q_i} \cdot  \frac{\mathrm{d}\vec m}{\mathrm{d} x_\mu} \right)   \mathrm{d}\vec{r}, \label{eq:Thiele:general:matrices} \\
\mathcal{G}_{i\mu}^\mathrm{SOT} &= -\int \vec m \cdot \left( \frac{\mathrm{d}\vec m}{\mathrm{d} q_i} \times \left(\vec m \times \hat{x}_\mu\right) \right)   \mathrm{d}\vec{r},  
&\mathcal{D}_{i\mu}^\mathrm{SOT} &= \int  \left(  \frac{\mathrm{d}\vec m}{\mathrm{d} q_i}  \cdot  \left(\vec m \times \hat{x}_\mu\right)  \right)   \mathrm{d}\vec{r}.  \nonumber
\end{align}
Here $x_\mu$ is the coordinate in the spatial direction $\mu$ and the corresponding unit vector is $\hat{x}_\mu$.
We would like to emphasise that the Thiele approach is only a good approximation if sufficiently many relevant coordinates are considered. 
Furthermore, it is only of practical quantitative use if the matrix elements can be computed with a reasonable effort, which can also involve numerical simulations.~\cite{Mueller2015} 

Note that for the above example of a translationally invariant system with a rigid magnetic texture with $\vec{q}=\vec{R}$ one obtains $\frac{\mathrm{d}\vec m}{\mathrm{d}\vec{R}_i} = - \frac{\mathrm{d}\vec m}{\mathrm{d}x_i}$.
Hence, the gyro-matrix $\mathcal{G}$ and the STT-coupling matrix $\mathcal{G}^\mathrm{STT}$ are directly related via $\mathcal{G}_{XY}=-\mathcal{G}_{Xy}^\mathrm{STT} = -4\pi\mathcal{Q}$, where $\mathcal{Q}$ is the skyrmion winding number, see Eq.~\eqref{eq:skyrmionnumberQ}.
Similarly, in this standard Thiele approach, the dissipation matrix $\mathcal{D}$ and the dissipative STT-coupling matrix $\mathcal{D}^\mathrm{STT}$ are related via $\mathcal{D}_{ij}=-\mathcal{D}_{ij}^\mathrm{STT}$ and their components resemble the magnetic stiffness in the energy functional, see Eq.~\eqref{eq:energy}.

In the following, we apply the generalized Thiele equations to describe the motion of magnetic solitons focusing on domain wall and skyrmion dynamics.

%%%%%%%%%%%%%%%%%%%%%%%%%%%%%%%%%%%%
%%%%%%%%%%%%%%%%%%%%%%%%%%%%%%%%%%%%
\subsection{Magnetization dynamics of domain walls in nanowires}

Magnetic domain walls can be moved by various sources, including, in particular, magnetic fields and spin-currents. 
The details of the motion as well as their possible maximal velocity typically depend on details of the system and the relevant magnetic interactions. 
In systems without DMI, for example, the plane in which the magnetization rotates when passing through the domain wall, i.e., \emph{domain wall angle} or \emph{helicity}, is determined by magnetostatic interactions, which are a rather weak effect.
When increasing the driving magnetic field above a certain threshold value, the helicity unpins and the magnetization inside the domain wall precesses. 
This effect, known as the \emph{Walker breakdown}~\cite{Schryer1974}, leads to a reduced domain wall speed and is therefore detrimental for the application in information technology, as discussed in Sec.~\ref{sec6:applications}.
Nowadays, it is possible to design materials which have a strong DMI that more strongly pins the helicity and, consequently, raises the barrier for the activation of the Walker breakdown.

The magnetic field-driven dynamics of one-dimensional magnetic domain walls have been extensively studied over many decades and can be well described in the Thiele framework.
Also magnetic domain walls in higher dimensions can be well described by this simple technique. 
Here, additionally to the one-dimensional case, the position of the domain wall is not only a one-dimensional parameter, but characterized by an extended line or surface. 
The additional degrees of freedom that then typically become relevant is the tilting~\cite{Malozemoff1979} or bending of the hyperplanes of the domain walls.

To demonstrate the Thiele approach, let us consider a domain wall in an effectively one-dimensional system. 
This means we assume that a domain wall is located in a nanowire which is narrow compared to the length scale of the variations of the magnetic texture. 
In such a system, a simple ansatz for the domain wall profile can be written as
\begin{equation}
\label{eq:dwansatz}
\vec{m}\left(x-X,\psi\right) = \left(\cos\psi\sin\theta\left(x-X\right), \sin\psi\sin\theta\left(x-X\right), \cos\theta \left(x-X\right)\right).
\end{equation}
where $X$ and $\psi$ are the position and the helicity of the domain wall, respectively, where $\psi=\pm\pi/2$ describes a Bloch type wall and $\psi=0$ or $\pi$ describes a N\'eel type wall.
$\theta(x)$ is the azimuthal angle of the magnetization.
Here we assumed that the nanowire is along the $\hat{x}$-direction and that the helicity is not spatially dependent.

Consider now the standard thin film setup as introduced in Sec.~\ref{sec2:spintorques}, where the DMI is the relevant source of the twisting of the magnetization and magnetostatic interactions only enters on the level of a modified uniaxial anisotropy.
In its simplified form, the only parameters that enter the energy functional Eq.~\eqref{eq:energy} for a low energy description are the uniform exchange $A$, interfacial DMI $D>0$, and the easy-axis anisotropy $K>0$.
In one spatial dimension, the energy functional then explicitly takes the form
\begin{equation}
 E[\vec{m}] = \int
    A\, \left(\frac{\mathrm{d}\vec{m}}{\mathrm{d}x}\right)^2
    - D\, \left( m_x \frac{\mathrm{d}m_z}{\mathrm{d}x} - m_z\frac{\mathrm{d}m_x}{\mathrm{d}x}  \right)
    - K\, m_z^2
    \,\,\mathrm{d}x\,\,.
\end{equation}
A domain wall which connects two polarized phases $\vec m (-\infty) = -\hat{z}$ and $\vec m (\infty) = \hat{z}$ minimizes this energy functional for the profile
\begin{equation}
\label{eq:dwprofile}
 \theta(x) = -2 \arctan\left(e^{-\sqrt{K/A} \, x}\right)
 \quad \text{and} \quad
 \psi = \pi \,\,.
\end{equation}
Here, the DMI term fixes the helicity $\psi = \pi$ while the other terms are independent of $\psi$.
In the following, we will use the ansatz, Eq.~\eqref{eq:dwansatz}, and the profile, Eq.~\eqref{eq:dwprofile}, to discuss the current-driven motion of domain walls on the Thiele level.

Note that, in broader nanowires, the additional spatial dimension can allow for more complex domain wall profiles and also dynamics. In particular, domain walls in finite-width systems with DMI show a tilting of the domain wall normal~\cite{Boulle2013} which can be explained by the interaction with the edges of the system~\cite{Muratov2017}. The dynamics of Bloch lines are also known to lead to more complex behavior.\cite{Malozemoff1973} However, these effects go beyond the scope of this introduction. 

\subsubsection{Domain wall motion due to spin-transfer torques}

In the continuum limit, without inhomogeneities the system is translationally invariant, i.e., $F_X=0$.
We will first consider the case of a small driving current which can only activate the zero mode, i.e., the translational mode.
Next we will discuss the case of stronger driving which leads to the activation of the helicity degree of freedom and, finally, to the Walker breakdown under STTs.

\begin{figure}[t]
\sidecaption[t] 
\includegraphics[width=0.64\columnwidth]{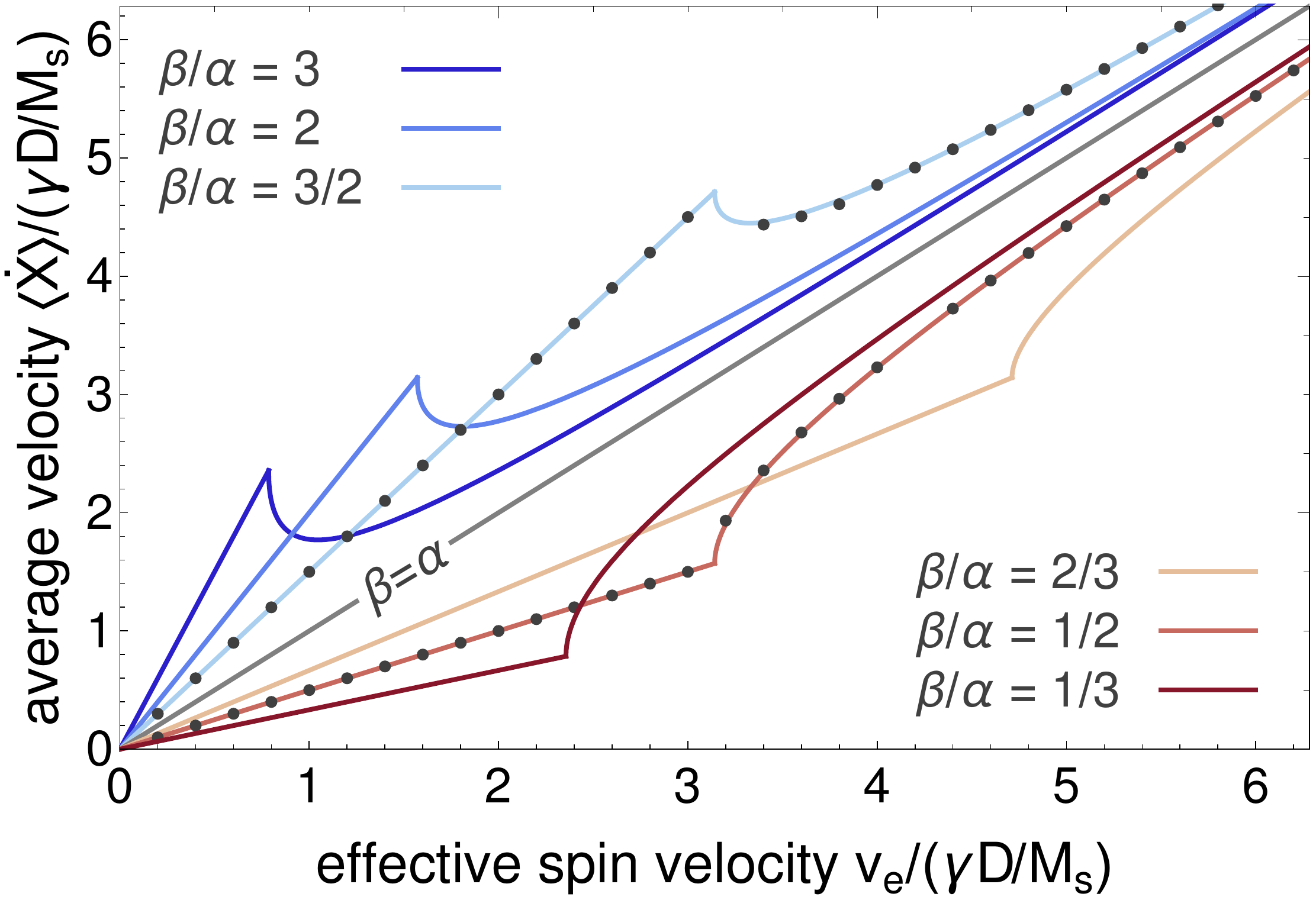}   
\caption{STT-driven domain wall motion.
    Shown is the average domain wall velocity $\langle \dot{X} \rangle$ as function of the effective spin velocity $v_e$.
    Solid lines show Thiele results, see Eq.~\eqref{eq:motion:dw:stt} and Eq.~\eqref{eq:motion:dw:sttwalker}, respectively. 
    The dots are LLGS simulation results, see Eq.~\eqref{eq:LLGS}.
    We fixed $\alpha=0.1$ for various $\beta$, i.e., $\beta=3\alpha, 2\alpha, \frac{3}{2}\alpha$ (dark to light blue), $\beta=\alpha$ (gray dashed), and $\beta=\frac{2}{3}\alpha, \frac{1}{2}\alpha, \frac{1}{3}\alpha$ (light to dark red).
}
\label{fig5thielewalker}   
\end{figure}

\runinhead{Pinned helicity.} 
In the limit of a small STT $\vec v_e = v_e \, \hat{\vec x}$, only the true zero modes are activated.
Therefore, for the one-dimensional domain wall, the only relevant collective coordinate is the position $X$.
Due to the lack of further spatial dimensions, the only terms which contribute from Eq.~\eqref{eq:Thiele:general} are the dissipation terms.
Since $\mathcal{D}_{XX}=-\mathcal{D}_{Xx}^\mathrm{STT}$, however, the Thiele equation reduces to the simple expression
\begin{equation}
 \label{eq:motion:dw:stt}
 \dot{X} = \frac{\beta}{\alpha} v_e .
\end{equation}
In the limit of a rigid texture, the velocity $\dot{X}$ is directly proportional to the effective spin velocity $v_e$, and it is completely independent of details of the domain wall shape, see Fig.~\ref{fig5thielewalker}.

\runinhead{Unpinned helicity.} 
In a next step, we consider the role of collective coordinates beyond the translational zero mode.
The position $X$ is still a zero mode with $F_X(X,\psi)=0$ and, moreover, the off-diagonal dissipation matrix elements vanish, i.e., $\mathcal{D}_{X\psi}=\mathcal{D}_{\psi x}^\mathrm{STT}=0$. 
Thus, the two coupled Thiele equations 
read
\begin{subequations}
\begin{alignat}{3}
\label{eq:Thiele:generaldwa}
    &\mathcal{G}_{X\psi} \dot{\psi}  
    &\ +\ & \mathcal{D}_{XX} (\alpha \dot{X} - \beta v_e) &\ =\ & 0 \,\,,
    \\
\label{eq:Thiele:generaldwb}
    &\mathcal{G}_{\psi X} (\dot{X} - v_e) 
    &\ +\ & \mathcal{D}_{\psi \psi} \, \alpha \, \dot{\psi}  &\ =\ & F_\psi(X,\psi) \,\,.
\end{alignat}
\end{subequations}
With the ansatz from Eq.~\eqref{eq:dwansatz} and the solution for the profile in Eq.~\eqref{eq:dwprofile}, the gyro-coupling and dissipation matrices of the Thiele equations evaluate to
\begin{subequations}
\begin{alignat}{3}
    &\mathcal{G}_{\psi X}= -\mathcal{G}_{X\psi}
    &\ =\ & 
    m_z(\infty)-m_z(-\infty) 
    &\ =\ & 
    2 \,\,,
    \\
    &\mathcal{D}_{XX}
    &\ =\ &
    \int_{-\infty}^\infty (\theta'(x))^2 \, \mathrm{d}x 
    &\ =\ &
    2\sqrt{K/A} \,\,,
    \\
    &\mathcal{D}_{\psi\psi}
    &\ =\ &
    \int_{-\infty}^\infty \sin^2\theta(x) \, \mathrm{d}x 
    &\ =\ &
    2\sqrt{A/K}\,\,.
\end{alignat}
\end{subequations}
For a non-equilibrium helicity, i.e.\ $\psi\neq\pi$, the DMI term yields a positive energy contribution while the other terms remain unaffected.
Relative to the energy of the domain wall in equilibrium, we therefore obtain the energy $E(X,\psi)$ and force $F_\psi(X,\psi)$
\begin{equation}
 E(X,\psi) = \pi D (1+\cos\psi) 
 \quad \Rightarrow \quad 
 F_\psi(X,\psi)=- \frac{ \gamma}{M_s}\partial_\psi E(\psi)= \frac{ \pi \gamma D}{M_s} \sin\psi
\end{equation}
which completes the constituents of Eqs.~\eqref{eq:Thiele:generaldwa} and \eqref{eq:Thiele:generaldwb}.
This set of coupled non-linear differential equations can be solved analytically, in both cases, (i) below and (ii) above the Walker-like breakdown.

\subruninhead{Below the Walker breakdown,} the helicity rotates away from its equilibrium position and, in the long-time limit, assumes a constant value, i.e., $\dot\psi=0$.
In this limit, Eq.~\eqref{eq:Thiele:generaldwa} reduces to the simplified case Eq.~\eqref{eq:motion:dw:stt} where the helicity dynamics are absent and the velocity $\dot{X}$ is independent of details of the domain wall texture.
From Eq.~\eqref{eq:Thiele:generaldwb}, we obtain the current-dependent helicity $\psi(v_e)$ of the driven domain wall which gives
\begin{equation}
 \psi(v_e) = \pi + \arcsin\left( \frac{\alpha-\beta}{\alpha} \frac{2 M_s v_e}{\pi \gamma D} \right) \quad \text{for }\  |v_e| \leq v_e^c = \frac{\alpha}{|\alpha-\beta|} \frac{\pi\gamma D}{2M_s}\,\,.
\end{equation}
For currents above the critical current $v_e^c$  the restoring force $F_\psi(X,\psi)$ cannot compensate for the velocity anymore and, therefore, solutions with $\dot\psi =0$ can no longer be obtained.
Consequently, $v_e^c$ marks the onset of the Walker breakdown.

\subruninhead{Above the Walker breakdown,} we can solve Eq.~\eqref{eq:Thiele:generaldwa} for $\dot X$ and make Eq.~\eqref{eq:Thiele:generaldwb} an equation of only $\psi$ and $\dot\psi$.
This differential equation can be solved exactly for a constant current density $v_e$ and the solutions can be written in the form 
\begin{subequations}
\begin{align} 
\label{eq:thieledw:abovewalkera}
\dot{X}(v_e,t) &= \frac{\beta}{\alpha} v_e + \frac{\sqrt{A/K}}{\alpha} \dot{\psi}(v_e,t) \,\,,\\
\label{eq:thieledw:abovewalkerb}
\psi(v_e,t)       &= -2\, \mathrm{arccot} \left( \frac{ u\,\,\mathrm{sign}(\alpha-\beta)}{1 - \sqrt{u^2-1} 
\tan( \omega_\psi \, t ) 
} 
\right)
\quad \text{with } u=\frac{v_e}{v_e^c}\geq1.
\end{align}
\end{subequations}
Here $T=\pi/\omega_\psi$ is the period of one helicity rotation and $\omega_\psi$ is the frequency given by
\begin{equation}
  \omega_\psi = \frac{\alpha}{1+\alpha^2} \frac{\pi\gamma D}{4 M_s} \sqrt{\frac{K}{A}} \sqrt{\Bigl(\frac{v_e}{v_e^c}\Bigr)^2-1} \,\,.
\end{equation}
As can be seen from Eq.~\eqref{eq:thieledw:abovewalkera}, the velocity $\dot X$ of the domain wall is also periodic with the frequency $\omega_\psi$ and shows a very complicated behavior as function of time.
The average velocity $\langle\dot X\rangle$, however, can be obtained from the time-average of Eq.~\eqref{eq:thieledw:abovewalkera} where we can exploit the simple relation $\langle\dot \psi\rangle = (2\pi/T)\,\mathrm{sign}(\alpha-\beta)$.
This yields
\begin{equation}
\label{eq:motion:dw:sttwalker}
\langle\dot X\rangle =  \frac{\beta}{\alpha}v_e + \frac{\mathrm{sign}(\alpha-\beta)}{1 + \alpha^2} \frac{\pi\gamma D}{2 M_s} \sqrt{\Bigl(\frac{v_e}{v_e^c}\Bigr)^2-1} 
\quad \text{for }  |v_e| \geq v_e^c
\end{equation}
as the average velocity of the domain wall above the Walker breakdown, $v_e>v_e^c$, in the Thiele framework.
Interestingly, it turns out that above the Walker breakdown, for $\beta<\alpha$ the domain wall speed does not get reduced but boosted instead. 
In Fig.~\ref{fig5thielewalker}, we illustrate these different behaviors obtained from the Thiele approach, see Eq.~\eqref{eq:motion:dw:stt} and Eq.~\eqref{eq:motion:dw:sttwalker}.
For comparison, we also show data obtained from numerical simulations of the full LLGS equation, Eq.~\eqref{eq:LLGS}.

Despite the Walker breakdown there are other interesting effects for field-driven domain walls and magnetic bubbles~\cite{Malozemoff1979,Slastikov2019}. 
For example, for a time-dependent current, the coupled dynamics of the position and helicity degree of freedom, Eqs.~\eqref{eq:Thiele:generaldwa} and~\eqref{eq:Thiele:generaldwb}, lead to an effective mass similar to the \emph{D\"oring mass}.~\cite{Hubert1975}

\subsubsection{Domain wall motion due to spin-orbit torques}

Spin-transfer torques act via gradients $\vec{v}\!_e \cdot \nabla$ only on local changes of the magnetization.
This is very different for SOTs being characterized by a spin polarization $\vec{\sigma}$, where $\vec\sigma$ couples explicitly to the local direction of the magnetization. 
Therefore, these can apply a torque also on a uniform magnetization and, moreover, induce a helicity-dependence of the total forces. 
Upon including the helicity degree of freedom $\psi$ as a collective coordinate for the description of the domain wall dynamics, and using the ansatz from Eq.~\eqref{eq:dwansatz}, we obtain the following Thiele equations for the SOT-driven domain wall 
\begin{subequations}
\begin{alignat}{4}
\label{eq:Thiele:generalSOTdwa}
    \mathcal{G}_{X\psi} \dot{\psi}  
    &\ +\ 
    &\tau_\mathrm{FL} \vec{\sigma}\!\cdot\!\vec{G}_X^\mathrm{SOT}(\psi)
    &\ +\ 
    &\alpha \mathcal{D}_{XX} \dot{X} 
    &\ +\ 
    &\tau_\mathrm{DL} \vec{\sigma}\!\cdot\!\vec{D}_X^\mathrm{SOT}(\psi) 
    &\ =\ 0 \,\,,
    \\
\label{eq:Thiele:generalSOTdwb}
    \mathcal{G}_{\psi X} \dot{X}
    &\ +\ 
    &\tau_\mathrm{FL} \vec{\sigma}\!\cdot\!\vec{G}_\psi^\mathrm{SOT}(\psi)
    &\ +\ 
    &\alpha\mathcal{D}_{\psi \psi} \dot{\psi} 
    &\ +\ 
    &\tau_\mathrm{DL} \vec{\sigma}\!\cdot\!\vec{D}_\psi^\mathrm{SOT}(\psi) 
    &\ =\ F_\psi(X,\psi) \,\,.
\end{alignat}
\end{subequations}
Here, for better readability, we have summarized the matrix products from Eq.~\eqref{eq:Thiele:general} into scalar products of $\vec\sigma$ with SOT gyro or dissipation vectors.
For the SOT-independent terms we refer to the previous section.
For the domain wall ansatz in Eq.~\eqref{eq:dwansatz} together with the solution in Eq.~\eqref{eq:dwprofile}, these SOT-specific vectors read
\begin{subequations}
\begin{alignat}{4}
    &\vec{G}_X^\mathrm{SOT}(\psi)
    &\ =\ 
    &\vec\psi \sin\theta + \hat z \cos\theta|_{\theta(-\infty)}^{\theta(+\infty)} 
    &\ =\ 
    && 2 \, \hat z &\,\,,
    \\
    &\vec{G}_\psi^\mathrm{SOT}(\psi)
    &\ =\ 
    &\int_{-\infty}^\infty - \,(\hat z \!\times\! \vec\psi) \sin\theta(x)  \, \mathrm{d}x 
    &\ =\ 
    && \pi\sqrt{A/K} \, (\hat z \!\times\!\vec\psi) &\,\,,
    \\
    &\vec{D}_X^\mathrm{SOT}(\psi)
    &\ =\ 
    &\int_{-\infty}^\infty (\hat z \!\times\! \vec\psi)\, \theta'(x) \, \mathrm{d}x 
    &\ =\ 
    && \pi \, (\hat z \!\times\! \vec\psi)  &\,\,,
    \\
    &\vec{D}_\psi^\mathrm{SOT}(\psi)
    &\ =\ 
    &\int_{-\infty}^\infty \vec\psi \cos\theta(x)\sin\theta(x) - \hat z \sin^2\theta(x) \, \mathrm{d}x 
    &\ =\ 
    && -2\sqrt{A/K} \, \hat z &\,\,,
\end{alignat}
\end{subequations}
where we defined the helicity vector $\vec\psi = (\cos\psi,\sin\psi,0)$ for a more compact notation.
For the standard setup with $\vec{j}\!_e = j_e \hat x$ and where the spin-polarization is determined by the spin Hall effect, i.e., $\vec\sigma = \hat z \times \vec{j}\!_e = j_e \hat y$, the contributions of gyro and dissipation vectors $\vec{G}_X^\mathrm{SOT}(\psi)$ and  $\vec{D}_\psi^\mathrm{SOT}(\psi)$ vanish.
Moreover, for a Bloch-type domain wall with $\psi = \pm \pi/2$ the other scalar products also vanish and, hence, the Bloch wall remains unaffected by the SOT.
In contrast, for a N\'eel type domain wall, the scalar products are maximized.
The N\'eel wall with $\psi=\pi$ then moves in the direction the current $\vec{j}\!_e$ with additional dynamics of $\psi$ whereas it moves in the opposite direction for $\psi=0$.
For additional information on SOT-induced dynamics we refer to Ref.~\cite{Manchon2019}.

%%%%%%%%%%%%%%%%%%%%%%%%%%%%%%%%%%%%
%%%%%%%%%%%%%%%%%%%%%%%%%%%%%%%%%%%%
\subsection{Magnetization dynamics of two-dimensional solitons}

Skyrmions and related magnetic textures, see Sec.~\ref{sec3:chiralmagnetictextures}, are not only thought to have an enhanced (topological) stability, but their non-zero winding number $\mathcal{Q}$, see Eq.~\eqref{eq:skyrmionnumberQ}, also leads to a gyromagnetic tensor element $\mathcal{G}_{XY}=-4\pi\mathcal{Q}$ in the Thiele equation, Eq.~\eqref{eq:Thiele:general}. 
This gyromagnetic coupling induces a force, similar to the Magnus force in Newtonian mechanics which acts on rotating bodies, leading to very particular dynamics. 
These include a response perpendicular to extrinsic forces and an intrinsic \emph{skyrmion Hall effect}, which we both discuss in the following.

As for domain walls in one spatial dimension, to demonstrate the Thiele approach, let us introduce an ansatz for a skyrmion-like profile in an out-of-plane polarized background of the form
\begin{equation}
 \vec{m}(\vec{r}-\vec{R},\psi) = \left(  \cos(\phi+\psi) \, \sin\theta,  \sin(\phi+\psi) \, \sin\theta, \cos\theta  \right),
 \label{eq:skyrmionDummyProfile}
\end{equation}
where $\phi=\phi(\vec{r}-\vec{R})$ sets the inplane magnetic profile and $\theta=\theta(\vec{r}-\vec{R})$ determines the $m_z$ profile.
$\vec R$ is the position of the skyrmion and $\psi$ is the helicity.
For circular skyrmions $\vec{R}$ is usually their center position, the profile only depends on the radial coordinate $\rho=|\vec r - \vec R|$ and $\phi$ depends only the axial coordinate $\chi$ of the cylindrical coordinate system centered at $\vec R$.
In this convention, the Bloch-type skyrmion shown in Fig.~\ref{fig1}(b) is described by $\phi=\chi$, $\psi=-\pi/2$, and $\theta=\theta(\rho)$ with $\theta(0)=\pi$ and $\theta(\infty)=0$.
The antiskyrmion in Fig.~\ref{fig1}(c) is described by $\phi=-\chi$, $\psi=\pi/3$.
Other skyrmion-like structures, e.g., higher order skyrmions can be described with $\mathcal{Q}=-N$ by setting $\phi=N\chi$, and the topologically trivial \emph{skyrmionium} is characterized by $\theta(0)=2\pi$ and $\theta(\infty)=0$.

%%%%%%%%%%%%%%%%%%%%%%%%%%%%%%%%%%%%
\subsubsection{Pinning and deformation}
At ultra low current densities all magnetic solitons are pinned by material defects.
For skyrmion lattices, it has been shown experimentally that the critical current density for depinning is very low.~\cite{Jonietz2010, Schulz2012}
Theoretically, the influence of disorder on skyrmion lattices was studied in various micromagnetic simulations.~\cite{Iwasaki2013b}
The micromagnetic results agree well with particle model simulations which are based on the Thiele equation of motion of skyrmions which interact with each other and with random pinning sites.~\cite{Reichhardt2015}
Also for isolated skyrmions in the presence of defects, the pinning, depinning, and motion has been studied experimentally~\cite{Litzius2020} and can be described in a generalized Thiele equation when taking deformations due to defects into account.~\cite{Mueller2015} 
For simplicity, however, we will neglect pinning effects in the following.

Deformations of moving solitons also occur in the absence of impurities, for example, due to internal dynamics, as has been shown already in early studies in magnetic bubble dynamics with Bloch lines.~\cite{Malozemoff1973}
For skyrmions, which do not have Bloch lines, deformations can also arise due to spin-torques.
In this case, the matrix elements of the Thiele equation, Eq.~\eqref{eq:Thiele:general:matrices}, become dependent on the current strength which leads to non-trivial corrections of the particle-like motion.~\cite{Masell2020}
These effects, as well as deformations due to thermal fluctuations, interactions with defects or other magnetic textures can induce an effective mass for two-dimensional solitons which might potentially be described by the broader term \emph{automotion}.~\cite{Malozemoff1973}
In the limit where skyrmions can be treated as rigid objects, i.e., when the bound state excitation gap of the skyrmion is large, deformation effects can be neglected, as we will assume in the following.

%%%%%%%%%%%%%%%%%%%%%%%%%%%%%%%%%%%%
\subsubsection{Skyrmion motion due to external forces}
\label{sec:motion:skyrmion:F}
Historically, before spin-torques became an active research field, the motion of magnetic bubbles was studied intensively, for example, with pulsed field gradients. 
It was found that the bubbles do not move along the direction of the external force, but along a deflected direction that depends on the winding number $\mathcal{Q}$.~\cite{Malozemoff1973} 
For both, skyrmions and magnetic bubbles, the side-drift response can be understood within the Thiele approach, Eq.~\eqref{eq:Thiele:general}.
Moreover, this effect occurs not only for field gradients but for all forces $\vec{F}(\vec{q})$ in the Thiele equation, e.g., due to field gradients or the interaction with defects and other magnetic structures.
It is also the source of the unusual Brownian motion of skyrmions which in two dimensions diffuse less if the Gilbert damping $\alpha$ is reduced.~\cite{Muller2017,Schuette2014,Zazvorka2019}
In the following, analog to the domain wall case, we will first discuss the limit of a pinned helicity and then consider what happens beyond this limit.

%%%%%%%%%%%%%%%%%%%%%%%%%%%%%%%%%%%%
\runinhead{Pinned helicity.}
Consider a skyrmion with winding number $\mathcal{Q}$ and  
the position $\vec R = (X,Y)$ 
as the only collective coordinates for the Thiele approach.
In a spatially dependent energy landscape $E(\vec R)$, e.g., due to anisotropy gradients, magnetic field gradients, defects, or other magnetic textures, $\vec R$ is not a true zero mode but can still be a good collective coordinate. 
As the system is dissipative, the skyrmion will at some point be trapped in a local minimum of $E(\vec R)$.
An elegant form of the resulting Thiele equation then reads
\begin{equation}
 \vec{G} \times \dot{\vec{R}} + \alpha \mathcal{D}(\vec R) \dot{\vec{R}} = \vec{F}(\vec{R}) 
\label{eq:Thiele:externalforceonly} 
\end{equation}
where $\vec{F}(\vec{R})  = -(\gamma/M_s)\nabla_{\vec{R}} E(\vec{R})$ is the force on the skyrmion, $\vec{G}=4\pi\mathcal{Q}\hat{z}$ is the gyro-vector, and $\mathcal{D}(\vec R)$ is the dissipation matrix.
The gyro-vector $\vec G$ couples the motion of the $X$ and $Y$ coordinates and leads to the side-deflection in the motion of two-dimensional solitons with a finite topological charge.
For a circular skyrmion, using the notation in Eq.~\eqref{eq:skyrmionDummyProfile} and the angular dependence $\phi(\chi)=N\chi$, $N\in\mathbb{Z}$, the dissipation matrix reduces to a scalar with
\begin{equation}
 \mathcal{D}(\vec R) = D_s(\vec R)\, \mathbb{1} = \mathbb{1}  \int_0^\infty \pi \rho \left( \frac{N^2}{\rho^2} \sin^2\theta(\rho) + \left( \theta'(\rho)\right)^2 \right) \, \mathrm{d}\rho 
\label{eq:Thiele:externalforceonly:D}
\end{equation}
where $\rho$ is the distance 
to $\vec{R}$, i.e., the center of the skyrmion.
Usually, it is assumed that the texture of the skyrmion does not change much with the position such that $D_s(\vec R)\approx D_s$ is a good approximation.

The Thiele equation, Eq.~\eqref{eq:Thiele:externalforceonly}, can be solved for the skyrmion velocity $\dot{\vec{R}}$.
Its absolute value $|\dot{\vec{R}}|$ and the direction relative to the force $\vec F$, parameterized by the deflection angle $\theta_\mathrm{d}$, then read
\begin{equation}
|\dot{\vec{R}}| = \frac{|\vec{F}(\vec R)|}{\sqrt{(4\pi\mathcal{Q})^2 + \alpha^2 D_s^2}} \quad\text{and}\quad  \theta_\mathrm{d} = -\arctan\left(\frac{4\pi\mathcal{Q}}{\alpha D_s}\right) \,\,.
\label{eq:Thiele:externalforceonly:rdotandtheta}
\end{equation}
In this formulation, the real-space topological nature of the side-deflection can be identified as $\theta_d\neq0$ only for $\mathcal{Q}\neq0$.
The deflection angle $\theta_d\neq0$ is schematically illustrated in Fig.~\ref{fig6skyrmionHallAngles} for various skyrmion-like textures and Gilbert dampings $\alpha$.
Moreover, Eq.~\eqref{eq:Thiele:externalforceonly:rdotandtheta} reveals that a finite charge $\mathcal{Q}$ reduces the (absolute) velocity $|\dot{\vec{R}}|$.
The dependence on $\mathcal{Q}$ should, however, be investigated more thoroughly as the dissipation scalar $D_s$ depends explicitly on the vorticity $N$, see Eq.~\eqref{eq:Thiele:externalforceonly:D}, and, hence, also on the winding number.
Furthermore, magnetic textures with different winding numbers usually relax to different magnetization profiles and, thus, different values of $D_s$.\cite{Malozemoff1973}

\runinhead{Unpinned helicity.}
Let us assume now that the skyrmion is stabilized in a system where the helicity $\psi$ is a zero mode and can, in principle, be activated.
However, this activation is not straightforward.
In the Thiele equation, Eq.~\eqref{eq:Thiele:general},  all matrix elements $\mathcal{G}_{X\psi}$, $\mathcal{G}_{Y\psi}$, $\mathcal{D}_{X\psi}$, and $\mathcal{D}_{Y\psi}$ vanish for circular solitons. 
Therefore, $\psi$ does not couple to the position $\vec R$ or derivatives thereof, which seemingly suggests that the helicity does not show any dynamics.
However, this conclusion is wrong as, for example, simulations with a magnetic field gradient show a steady rotation of the helicity while the skyrmion moves towards the direction of the smaller field.~\cite{Liang2018,Lohani2019}  
In the following, we discuss this example in more detail and show how to resolve the apparent contradiction.

Consider a magnetic field of the form $\vec{B}(\vec{r}) = (B_0 + x \,\delta B) \, \hat z$.
Let us assume, moreover, that the field gradient  $\delta B$ is a sufficiently small so that the skyrmion profile is still approximately circular and the above arguments still hold.
Due to the field gradient, the position $\vec R$ is not a zero mode but still a good collective coordinate which is subject to a force $\vec{F}\!_{\vec R}$ which drags the skyrmion towards regions with lower field.
While moving there, however, the skyrmion profile has to adapt to the local magnetic field $\vec{B}(\vec{r})$, leading to an inflation of the skyrmion size $\xi$.
Unlike $\vec R$, the collective coordinate $\xi$ couples directly to the helicity $\psi$ via the matrices in the Thiele equation but, due to the circular shape, $\xi$ does not couple to $\vec R$. 
In a compact form, the four Thiele equations then read
\begin{subequations}
\begin{alignat}{3}
    \label{eq:Thiele:externalforceonly:unpinned:thieleA}
    &\vec{G} \times \dot{\vec{R}} \
    & + \ & \alpha D_s \dot{\vec{R}}
    &=\ &  \vec{F}\!_{\vec R}(\vec R, \xi) \,\,,
    \\
    \label{eq:Thiele:externalforceonly:unpinned:thieleB}
    &\mathcal{G}_{\psi\xi} \ \dot{\xi}  \
    & + \ & \alpha \mathcal{D}_{\psi\psi} \ \dot{\psi}\ 
    &=\ & 0 \,\,,
    \\
    \label{eq:Thiele:externalforceonly:unpinned:thieleC}
    &\mathcal{G}_{\xi\psi} \ \dot{\psi}\
    & + \ & \alpha\mathcal{D}_{\xi\xi} \ \dot{\xi}
    &=\ & F_\xi(\vec R, \xi) \,\,.
\end{alignat}
\end{subequations}
All matrix elements with indices $\psi$ or $\xi$ are, in principle, dependent on $\xi$.
This dependence can be neglected on small time scales.
The force $F_\xi(\vec R, \xi)$ ensures that the skyrmion size adapts to the local magnetic field.
In a small field gradient $\delta B$, the skyrmion moves slow enough that we can assume $\xi$ to be close to the energetically optimal value.
Then its contribution to the force $\vec{F}\!_{\vec R,\xi}$ can be neglected and, to lowest order in $\delta B$, this force is $\vec{F}\!_{\vec R}\propto -\delta B\,\hat x$.
Now, Eq.~\eqref{eq:Thiele:externalforceonly:unpinned:thieleA} is decoupled from the other two equations of motion and the skyrmion moves according to the results of the previous section, Eq.~\eqref{eq:Thiele:externalforceonly:rdotandtheta}.
In particular, the parallel velocity is $\dot X \propto -\delta B$ and, therefore, $\dot\xi \propto \dot B(\vec R) = \dot X \delta B \propto \delta B^2$.
Eq.~\eqref{eq:Thiele:externalforceonly:unpinned:thieleB} then yields the velocity of the helicity $\dot\psi \propto \delta B^2/\alpha$ which continuously rotates while the skyrmion moves in magnetic field gradient,~\cite{Liang2018,Lohani2019}  similar to the domain wall above the Walker breakdown.~\cite{Schryer1974} 

We would like to point out that $\dot\psi\propto \delta B^2/\alpha$ is also the consequence of another effect which we did not capture in the above discussion:
So far, we assumed that the skyrmion maintains its circular shape.
In the field gradient $\vec B(\vec R)$, however, the skyrmion becomes slightly non-circular which adds a finite direct coupling $\mathcal{D}_{X\psi} \propto \delta B$ between the velocity $\dot X$ and $\dot\psi$ to Eqs.\eqref{eq:Thiele:externalforceonly:unpinned:thieleA} and \eqref{eq:Thiele:externalforceonly:unpinned:thieleB}.
For the full dynamics, therefore, both the change of the skyrmion size and the its non-circular distortion contribute. 

In the following sections, we review the dynamics of current-driven instead of force-driven skyrmions which follow the same basic concepts.

%%%%%%%%%%%%%%%%%%%%%%%%%%%%%%%%%%%%
\subsubsection{Skyrmion motion due spin-transfer torques}
\label{sec:motion:skyrmion:STT}
A standard example for the application of a Thiele equation is to model the motion of spin-tranfer torque-driven skyrmions in chiral magnets.
The Thiele formalism, for example, provides a direct mean to explain the \emph{skyrmion Hall effect}, where the skyrmions move at an angle relative to the direction of the applied current density $\vec{j}\!_e$. We will discuss this in the following. 
Analog to the previous chapters we will first discuss the limit of a pinned helicity and then explain briefly what happens beyond this regime.

\runinhead{Pinned helicity.}
Let us consider a frequently used assumption for skyrmions in chiral magnets, namely that the helicity is pinned by DMI to a fixed value $\psi$ and does not contribute to the dynamics.
Consider, moreover, that the system is translation invariant, i.e., the position $\vec R$ is a zero mode, and that the skyrmion can be described by the ansatz in Eq.~\eqref{eq:skyrmionDummyProfile}.
The Thiele equation then reads
\begin{equation}
 \vec{G} \times (\dot{\vec{R}}-\vec{v}\!_e) + D_s (\alpha\dot{\vec{R}}-\beta\vec{v}\!_e) = 0
 \label{eq:Thiele:skyrmion:STT}
\end{equation}
where $\vec{G}=4\pi\mathcal{Q}\hat{z}$ is the gyro-vector, and the dissipation matrix $\mathcal{D}$ reduces to a scalar $D_s$ as in Eq.~\eqref{eq:Thiele:externalforceonly:D}.
In principle, this equation of motion can be solved for $\dot{\vec R}$ which yields the skyrmion Hall effect.
Alternatively, we can interpret the effect of the STTs from a different perspective.
By isolating all terms which originate from the STT on the right hand side of Eq.~\eqref{eq:Thiele:skyrmion:STT}, we effectively recover the Thiele equation for a skyrmion driven by an external force, Eq.~\eqref{eq:Thiele:externalforceonly}, with
\begin{equation}
 \vec{F}^\mathrm{STT} = \vec{G} \times \vec{v}\!_e + \beta D_s \vec{v}\!_e \,\,. 
\end{equation}
The skyrmion Hall angle $\theta_d^\mathrm{STT}$ is then the sum of (i) angle $\theta_F^\mathrm{STT}$ between the effective STT-force $\vec{F}^\mathrm{STT}$ and the direction of the current $\vec{v}\!_e$ and (ii) the deflection angle $\theta_d$ for a force-driven skyrmion, see Eq.~\eqref{eq:Thiele:externalforceonly:rdotandtheta}, and reads
\begin{equation}
 \label{eq:Thiele:skyrmion:STT:Hall}
 \theta_d^\mathrm{STT} 
 = \arctan\left(\frac{4\pi\mathcal{Q}}{\beta D_s}\right) - \arctan\left(\frac{4\pi\mathcal{Q}}{\alpha D_s}\right) 
 =  \arctan\left(\frac{4\pi\mathcal{Q} D_s (\alpha-\beta)}{ (4\pi\mathcal{Q})^2 + \alpha\beta D_s^2}\right) \,\,.
\end{equation}
The result reflects the trivial cases $\theta_d^\mathrm{STT}=0$ for $\alpha=\beta$ and for $\mathcal{Q}=0$ where the magnetic texture just moves along with the current.
In contrast to the deflection angle $\theta_d$ in the previous section, Eq.~\eqref{eq:Thiele:externalforceonly:rdotandtheta}, the skyrmion Hall angle $\theta_d^\mathrm{STT}$ shrinks for increasing $\mathcal{Q}$ and, for typical values of parameters, the maximal $\theta_d^\mathrm{STT}$ is at $\mathcal{Q}=\pm1$. 
The properties of $\theta_d^\mathrm{STT}$ for different skyrmion-like solitons are schematically summarized in Fig.~\ref{fig6skyrmionHallAngles}.

The skyrmion Hall angle can change dramatically, for example, when considering small random defects which additionally reduce the velocity.~\cite{Mueller2015,Reichhardt2015} 
Extended defects, such as the DMI-induced twisting at the edge of a sample, in turn, can speed up the skyrmion motion $\propto |\mathcal{Q}|/\alpha$ when STTs push the skyrmion into nonequilibrium positions.~\cite{Iwasaki2013b}
To accommodate the translationally non-invariant case in the Thiele formalism one has to take a spatially dependent force in Eq.~\eqref{eq:Thiele:skyrmion:STT} into account.

Moreover, the STT-induced torques can distort the skyrmion profile on the level of the LLGS equation which eventually leads to strong corrections to the skyrmion Hall effect and, even more importantly, a speed limit above which the STTs destroy the skyrmion.~\cite{Masell2020}
The latter cannot be derived from a simple Thiele ansatz and requires more rigorous models or numerical simulations of the LLGS equation.

\runinhead{Unpinned helicity.}
As discussed above in Sec.~\ref{sec:motion:skyrmion:F}, for a circular skyrmion, the coupling between the collective coordinates $\vec R$ and the helicity $\psi$ is absent.
Similarly, because of $\partial_X \vec m = - \partial_x \vec m$, STTs do not directly couple to the helicity.
However, STT-driven skyrmions can still show dynamics of the helicity, e.g., in an energy landscape $E(\vec R, \psi)$ where the position and helicity are coupled.
In this case, the helicity of the moving skyrmion shows dynamics around the local optimum $\psi_0(\vec R)$, potentially showing features of an effective helicity mass.~\cite{Leonov2016c}
Moreover, STTs can deform the skyrmion which enables the coupling in the Thiele equation, leading to a steady rotation of the helicity.

\subsubsection{Skyrmion motion due spin-orbit torques}
\label{sec:motion:skyrmion:SOT}
In contrast to STTs, SOTs couple directly to the magnetization texture and not derivates thereof.
Therefore, SOT induced dynamics are sensitive to the helicity $\psi$ of the magnetic soliton as we will discuss in more detail in the following.
A skyrmion Hall effect can also be derived for skyrmions with SOTs and was recently also confirmed experimentally~\cite{Litzius2020,Litzius2017}.

\runinhead{Pinned helicity.}   
In monolayers on heavy metal substrates or thin films of stacked heterostructures, skyrmions can be stabilized by a strong DMI with extra stabilizing support from dipolar interactions.
These interactions usually pin the helicity such that it does not contribute to the dynamics.

For the axially symmetric soliton with $\vec{m}(\infty)=\hat{z}$ and $\phi=N\chi$ in the ansatz in Eq.~\eqref{eq:skyrmionDummyProfile}, the gyro-coupling SOT matrix evalutes to zero whereas the dissipative SOT coupling matrix elements only vanish for $|N|\neq1$.
Note that in this convention, the relation $\mathcal{Q} = N(1-m_z(0))/2$ implies that not only the skyrmion and the antiskyrmion, but also the topologically trivial skyrmionium can be driven by SOTs.
In turn, higher order skyrmions with $|\mathcal{Q}|>1$ do not react to SOTs within these approximations.
The Thiele equation for these $|N|\neq1$ objects can be written as
\begin{equation}
 \vec{G} \times \dot{\vec{R}} 
 + \alpha D_s \dot{\vec{R}} 
  + \tau_\mathrm{DL} \, \vec\sigma\cdot\mathcal{D}_{\vec{R}}^\mathrm{SOT}(\psi)
  = 0 \,\,.
 \label{eq:Thiele:skyrmion:SOT}
\end{equation}
Here, we have introduced the $3\times2$ dissipation tensor $(\mathcal{D}_{\vec{R}}^\mathrm{SOT})_{\mu i}$, $\mu=x,y,z$, $i=X,Y$, for the SOT-induces torque which reads
\begin{equation}
\label{eq:Thiele:skyrmion:SOT:DSOT}
    \mathcal{D}_{\vec{R}}^\mathrm{SOT}(\psi)    =     \Big( 
    \hat z \times\vec\psi,
    -N \vec\psi \Big)
    \pi \delta_{|N|,1} \!\int_0^\infty\!\!  \cos\theta(\rho) \sin\theta(\rho) + \rho \theta'(\rho) \, \mathrm{d}\rho  
\end{equation}
with $\vec\psi=(\cos\psi,\sin\psi,0)$.
The Kronecker delta $\delta_{|N|,1}$ indicates that only solitons with $N=\pm1$ give finite contributions.
The asymmetric $N$-dependence in only the second column of $\mathcal{D}_{\vec{R}}^\mathrm{SOT}$ is an artefact of the ansatz, Eq.~\eqref{eq:skyrmionDummyProfile}, which flips $m_y\to-m_y$ for $N\to-N$.

\begin{figure}[th!]
\includegraphics[width=0.99\columnwidth]{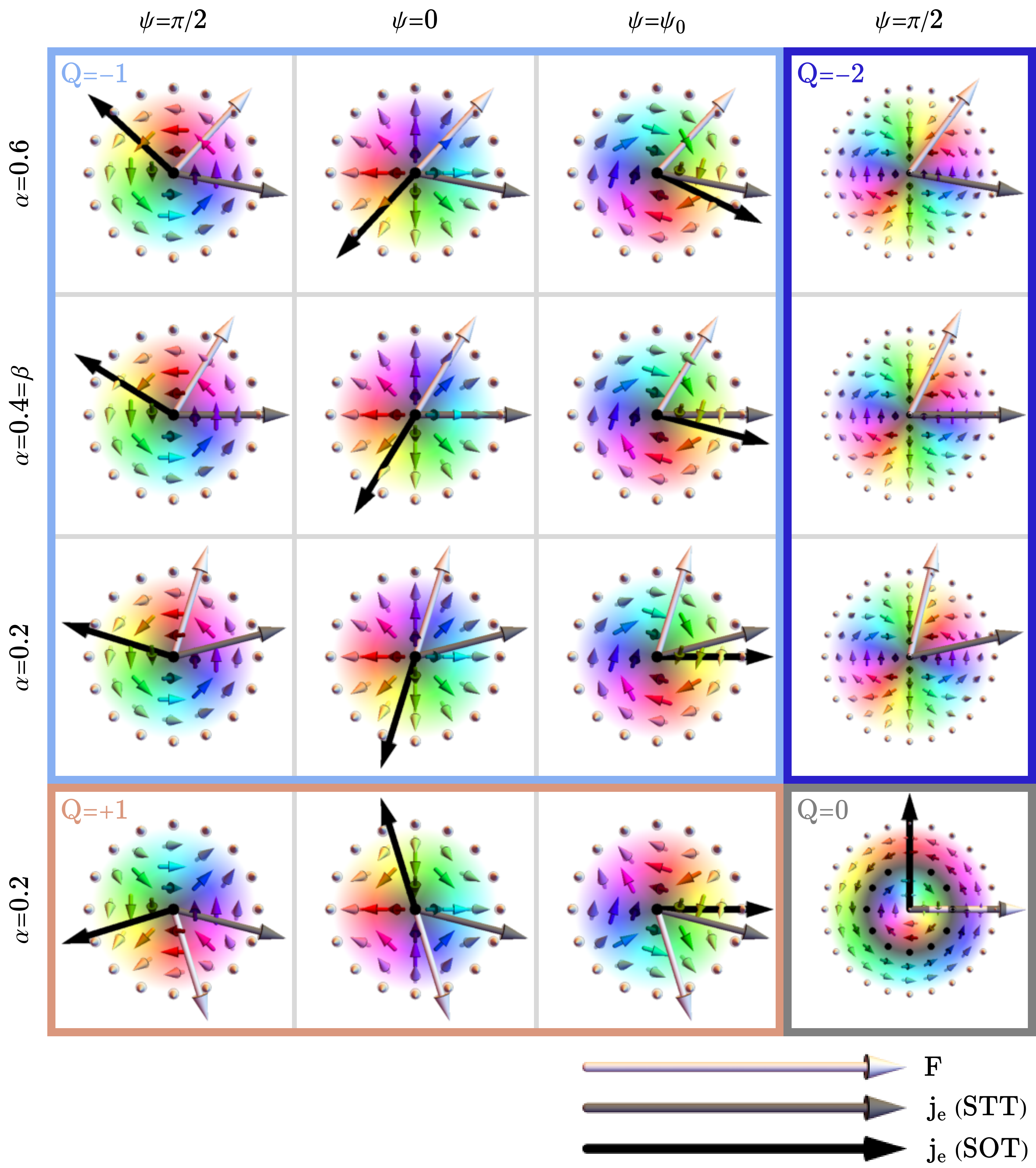}
\caption{
    Deflection angle for force-driven and skyrmion Hall angles for STT-driven and SOT-driven two-dimensional solitons.
    The force $\vec F = - (\gamma/M_s)\partial_{\vec R} E$ and the electric currents $\vec{j}\!_e$ for both STTs and SOTs point to the right, as indicated.
    For STTs we use $\beta=0.4$ in all panels and for SOTs we assume $\vec\sigma = \hat z \times \vec{j}\!_e$.
    The direction of the velocity of the reacting soliton is indicated by arrows in each panel, illustrating the Thiele results of Eqs.~\eqref{eq:Thiele:externalforceonly:rdotandtheta}, \eqref{eq:Thiele:skyrmion:STT:Hall}, and \eqref{eq:motion:skyrmion:SOT:Hall}.
    A skyrmion with $\mathcal{Q}=-1$ and an antiskyrmion with $\mathcal{Q}=1$ are shown, both for three different helicities $\psi=\pi/2$ , $0$, and $\psi_0$.
    The compensation helicity $\psi_0$ is chosen such that the skyrmion Hall angle with SOT vanishes for $\alpha=0.2$.
    The impact of the damping $\alpha$ is shown by various $\alpha=0.6$, $0.4$, and $0.2$, where the skyrmion Hall angle in the second row vanishes ($\alpha=\beta$).
    Moreover, we show a higher order $\mathcal{Q}=-2$ skyrmion and a $\mathcal{Q}=0$ skyrmionium, both with $\psi=\pi/2$.
    The $\mathcal{Q}=-2$ skyrmion is unaffected by SOTs and has slightly modified responses to forces and STTs, compared to the skyrmion with $\mathcal{Q}=-1$.
    The skyrmionium moves precisely in the direction of the force and STT, while its reaction to SOTs is solely determined by its helicity.
    }
\label{fig6skyrmionHallAngles} 
\end{figure}

Similar to the discussion in Sec.~\ref{sec:motion:skyrmion:STT}, we can interpret the SOT-induced terms as an external force 
$\vec{F}^\mathrm{SOT}$ and derive the skyrmion Hall angle from the direction of this effective force.
Assuming again the standard spin Hall setup for the SOT with $\vec{j}\!_e = j_e \hat x$ and $\vec\sigma = \hat z \times \vec{j}\!_e = j_e \hat y$, the skyrmion Hall angle $\theta_d^\mathrm{SOT}$ becomes
\begin{equation}
\label{eq:motion:skyrmion:SOT:Hall}
 \theta_d^\mathrm{SOT}(\psi) = 
 - N \left( \psi+\pi - \arctan\left(\frac{4\pi\mathcal{Q}}{\alpha D_s}\right) \right) 
 \quad\text{with}\quad
 |N|=1\,\,,
\end{equation}
which is only well-defined for $|N|=1$ as otherwise the soliton does not move.
Note that the skyrmion Hall angle $\theta_d^\mathrm{SOT}$ is a function of the helicity $\psi$ and can result in a motion in arbitrary directions, including parallel to the current, by fine-tuning the DMI.~\cite{Kim2018}
This angular dependence is also schematically summarized in Fig.~\ref{fig6skyrmionHallAngles} for various soliton configurations and parameters.

\runinhead{Unpinned helicity.}
For circular two-dimensional solitons driven by SOTs, the same  physics arises as for the other driving mechanisms, namely that neither the collective coordinates $\vec R$ couple directly to $\psi$ nor do the Thiele matrices $\mathcal{G}_{\psi i}^\mathrm{SOT}$ and $\mathcal{D}_{\psi i}^\mathrm{SOT}$ yield a finite coupling between $\vec\sigma$ and $\psi$ (except for $\mathcal{D}_{\psi z}^\mathrm{SOT}$, which is usually not relevant as $\sigma_z=0$).

A distinguished feature of SOTs is that they tilt the background magnetization. This naturally leads to deformations of the soliton, breaking the axial symmetry and enabling a finite coupling of $\psi$ and $\dot{\vec R}$, see Sec.~\ref{sec:motion:skyrmion:F}.
Thus, while moving with velocity $\dot{\vec R}$ at a skyrmion Hall angle $\theta_d^\mathrm{SOT}(\psi)$, see Eq.~\eqref{eq:motion:skyrmion:SOT:Hall}, the helicity $\psi$ changes which feeds back on $\theta_d^\mathrm{SOT}(\psi)$.  
Consequently, the skyrmion with an activated helicity degree of freedom can end up orbiting around a fixed point~\cite{Zhang2017} or, for sufficiently asymmetric energy landscapes $E(\psi)$, perform a trochoidal motion~\cite{Ritzmann2018} which is a combination of translation and orbiting.
Moreover, once the helicity becomes dynamical, it can also lead to an effective mass in the Thiele equation.~\cite{Leonov2016c,Zhang2017}

\subsection{Magnetization dynamics of three-dimensional hopfions}
As magnetic hopfions in chiral magnets have only recently been proposed theoretically, see Sec.~\ref{sec3:chiralmagnetictextures}, their dynamics are a field that is still much under investigation.

In thin films of chiral magnets with perpendicularly magnetized surfaces, hopfions are predicted to be stabilized due to geometric confinement.
The magnetic texture of the hopfion is then fixed by the DMI such that only translational modes can be activated easily.
For such a setup, it was shown theoretically that the STT-driven $H=1$ hopfion behaves like a skyrmionium, i.e., it moves like a two-dimensional soliton, straight along the applied current without any Hall angle.~\cite{Wang2019}

More complex dynamics are predicted for three-dimensional frustrated magnets: 
Here, the translation in all spatial dimensions and rotation around all axes are zero modes.
It was shown in a theoretical study by Liu \textit{et al.} \cite{Liu2020} that the STT-driven $H=1$ hopfion indeed rotates while moving with the current, adjusting such that its skyrmionium-like cross-section aligns perpendicular to the current.
Moreover, inside the hopfion, regions with positive and negative skyrmion charge $\mathcal{Q}$ are present which are subject to opposite skyrmion Hall angles.
As a consequence, the STT-driven hopfion either inflates or deflates, dependent on the direction of the current.
For a detailed description of the dynamics, featuring also a discussion in the Thiele framework, we refer to Ref.~\cite{Liu2020}.

%%%%%%%%%%%%%%%%%%%%%%%%%%%%%%%%%%%%
%%%%%%%%%%%%%%%%%%%%%%%%%%%%%%%%%%%%
%%%%%%%%%%%%%%%%%%%%%%%%%%%%%%%%%%%%
\section{Potential Applications}
\label{sec6:applications}

Based on the very rich playground of spintronics with chiral magnetic structures, several potential application have been proposed over the recent years. In the following we will briefly introduce some of them.

\subsection{Storage and logic technologies}

\runinhead{Magnetic racetrack.} 
The central idea behind the racetrack is that information is encoded by magnetic bits which are placed in a one-dimensional shift register device. 
Data can be accessed or written at a particular point of the nanowire. 
It has the great advantage that instead of moving mechanical parts, only the magnetic bits are moved, e.g., by spin-currents. 
In the classically suggested version~\cite{Parkin2008, Hayashi2008a}, the bits are magnetic domains, separated by domain walls. 
For the racetrack based on magnetic skyrmions~\cite{Fert2013}, the state of a bit can be represented by the presence or absence of a skyrmion. 
The latter has the advantage to circumvent the impact of edge roughness in the nanowire, as skyrmions opposed to domain walls do not touch the edge. 
However, it has also some disadvantages. 
In particular, the skyrmion Hall effect hinders the straight motion of the skyrmion through the nanowire.
To enhance the speed of the magnetic data, (synthetic) antiferromagnetic instead of ferromagnetic materials have been studied within the recent years. 
An antiferromagnetic coupling would also resolve the problem with the skyrmion Hall effect as in this case the forces in the direction perpendicular to the current direction cancel. 
Moreover, similar devices with closely packed skyrmions or other similar solitons have been suggested as the information encoded in not well-defined inter-skyrmion distances is very fragile.~\cite{Muller2017b}

\runinhead{Bubble memory.} 
In the 1970s and 1980s, before magnetic racetracks were discussed, memory devices exploiting magnetic bubbles have been commercially available. 
These are non-volatile two-dimensional shift register memories that exploit the magnetic field-driven motion of small magnetized areas -- the bubbles.~\cite{Bobeck1975, ODell1986}  

\runinhead{Magnetic transistor.}
Transistors as key elements for controlling integrated circuits and logic devices have also been proposed to be implemented based on magnetic textures, such as a domain wall based transistor~\cite{Ma2018c} or a skyrmion based transistor~\cite{Zhang2015g}
These exploit the gate-voltage controlled motion of the magnetic nanostructures. 

\runinhead{Magnetic logic.} 
Another key field in spintronics is the idea to create magnetic-based logic gates.~\cite{Allwood2005}  
This is, on the one hand, done by studying nano-magnetic logic, where nano-magnetic islands with a uniaxial-anisotropy represent the zero and one state based on their orientation with respect to this anisotropy direction, e.g., ``up'' and ``down''.
The other idea is to send signals through an appropriately shaped device, which represent the logical gates. 
This includes magnonic logic~\cite{Wagner2016} as well as logic based on chiral magnetic states such as skyrmions. 
An example of the latter was suggested by Zhang \textit{et al.} \cite{Zhang2015} which exploits the possibility to convert spin-torque driven skyrmion into domain walls in narrow wires.
In a convention where a logical 1 or 0 is represented by the presence or absence of a skyrmion, respectively, an OR gate and AND gate have been simulated by properly designing the width of the narrow wires.

\runinhead{Magnetic nano-oscillators.} 
Oscillators exploit the system's natural time scale and responses to external sources to provide a tunable frequency source. 
In magnetic texture based systems, these oscillators are naturally on the nano-scale and exploit, for example, the current-driven self oscillation of domain walls~\cite{Sharma2015} or skyrmions~\cite{Garcia-Sanchez2016}. 
While for ferromagnets the frequencies are in the GHz regime and can be tuned, e.g., by an externally applied magnetic field, they can be in the THz regime for antiferromagnetic materials, thereby bridging the THz gap.

\subsection{Unconventional spintronics-based computing schemes}
Within the recent years, more and more unconventional computational paradigms are being explored. 
Based on their low-energy consumption, compact nanometer size scale, and manipulability, magnetic textures could play an important role in the development of such novel computational technologies.~\cite{Finocchio2019, Grollier2020}  

\runinhead{Magnetic artificial neural networks.}
The vast progress within the field of artificial intelligence is mainly based on the widely enhanced available hardware power, while most of the concepts have been suggested already a few years ago. 
So as with deep artificial neural networks, which nowadays are widely used for different types of AI applications. 
However, so far they are mostly performed on the existing hardware which, due to the classical segmentation in computational units and storage, are not optimally suited for these types of applications as their power consumption shows.
Instead, alternative architectures which adjust to the deep neural network structure are proposed, with a focus of creating their central components, i.e., artificial synapses and neurons, in hardware.
There are also several suggestions for magnetic neuromorphic computing~\cite{Grollier2020, Li2017c} and how to implement artificial \emph{neurons}~\cite{Sengupta2016} and \emph{synapses}.~\cite{Sharad2012,Song2020} 
In particular, \emph{memristors},~\cite{Wang2009} i.e., devices whose resistance depends on the previous state, are suggested to function as a basis for synaptic applications. 
 
\runinhead{Spintronics based reservoir computing.}
Reservoir computing has the goal to exploit the response of a reservoir to simplify, for example, spatial-temporal recognition tasks. 
The reservoir itself projects the input into a higher dimensional space, where it becomes easier to classify.
For this concept to work, the reservoir needs to be a non-linear, complex reservoir with a short-term memory, which is fulfilled by several physical systems opening up the path for in-materio computing.~\cite{Appeltant2011} 
As spintronics systems often naturally fulfill these criteria for the reservoir and additionally provide a lot of tune-ability as well as complexity, together with their low energy consumption, they do provide a promising hardware-based solution for reservoir computing.~\cite{Tanaka2019}
It has been proposed that skyrmion fabrics are very well suited for reservoir computing applications.~\cite{Prychynenko2017} 

\runinhead{Stochastic computing.}
The ansatz of stochastic computing is to trade speed for accuracy, exploiting the law of large numbers where upon enhancing the number of experiments the result converges to the expection value. 
For example, one can stochastically multiply two numbers in-between zero and one, when interpreting them as a probability of having a one in a bit-string. 
For uncorrelated bit-strings the multiplication of these two numbers can then be efficiently calculated as sending the two bit-strings through an AND gate.
Spintronics offers a potential ansatz with respect to stochastic computing, as spintronics systems can naturally exhibit stochastic behavior.
Furthermore, recently a device which allows to reshuffle bit-strings based on magnetic skyrmions has been realized.~\cite{Pinna2018, Zazvorka2019} 
Such a skyrmion reshuffler allows to restore the decoherence between signals which possibly synchronized. 
A similar suggestion is to encode the information in probabilistic bits, also called p-bits. 
These are bits that fluctuate between 0 and 1 and, in this sense, interpolate between a classical bit and a q-bit. 
It has been suggested that magnetic states naturally provide a realization for such p-bits.~\cite{Camsari2019}

\runinhead{Topological quantum computing.}
Even more exotically, chiral magnetic states could contribute to topological quantum computing.  
It has been suggested that Majorana modes localize at skyrmions~\cite{Yang2016d} or compound structures of superconducting vortices and skyrmions.\cite{Hals2016,Nothhelfer2019,Rex2019}  
This might provide a path to perform the key operation of topological quantum computing, i.e., braiding of the localized modes with a non-Abelian statistics, via the manipulation of magnetic textures.

%%%%%%%%%%%%%%%%%%%%%%%%%%%%%%%%%%%%
%%%%%%%%%%%%%%%%%%%%%%%%%%%%%%%%%%%%
%%%%%%%%%%%%%%%%%%%%%%%%%%%%%%%%%%%%
\section{Conclusion}
\label{sec7:conclusion}
This book chapter presented an introduction to current-induced dynamics of chiral magnetic structures.
We briefly summarized the basic concepts for deriving a continuum theory of magnetization dynamics in Sec.~\ref{sec2:spintorques} and introduced \emph{domain walls}, \emph{(anti-)skyrmions}, and \emph{hopfions} as examples for magnetic solitons in Sec.~\ref{sec3:chiralmagnetictextures}.
In the main part of this chapter, we focused on the manipulation of magnetic textures by spin-troques, both due to spin-transfer and spin-orbit mechanisms.
We reviewed (i) selected creation processes for domain walls and skyrmions in Sec.~\ref{sec4:creation} and (ii) the motion of the above solitons in Sec.~\ref{sec5:motion} with a particular focus on the generalized Thiele method.
Finally, in Sec.~\ref{sec6:applications} we summarized already implemented or theoretically suggested applications of magnetic textures which are manipulated by electric currents.

The field of \emph{spintronics}, which explores the interplay of electric currents and the magnetization, has shown an enormous theoretical and experimental progress in the past years and a vast variety of possible new routes have emerged, including antiferromagnetic materials which are not discussed in this chapter.
We can look forward with excitement to the future of current-induced magnetization dynamics, what new physics and which new quasi-particles they will reveal in the future, and how they might eventually contribute to our everyday life.

\begin{acknowledgement}
JM is supported as an Alexander von Humboldt/JSPS International Research Fellow (Project No. 19F19815).
KES acknowledges funding from the German Research Foundation (DFG) Project No. EV 196/2-1, No. EV196/5-1, TRR 173 -- 268565370 (project B12), as well as from the Emergent AI Center funded by the Carl-Zeiss-Stiftung.
\end{acknowledgement}

\end{document}